\documentclass{elsarticle}
\usepackage{hyperref}
\usepackage{amsmath}
\usepackage{amsfonts}
\usepackage{amssymb}
\usepackage{bm}
\usepackage{graphicx}
\usepackage{epstopdf}
\DeclareMathOperator{\Tr}{Tr}

\newcounter{bla}

\journal{Computer Physics Communications}

\begin{document}
\mathchardef\mhyphen="2D

\begin{frontmatter}



\title{ComDMFT: a Massively Parallel Computer Package for the Electronic Structure of Correlated-Electron Systems}

\author[a]{Sangkook Choi\corref{author}}
\author[a]{Patrick Semon}
\author[a]{Byungkyun Kang}
\author[a]{Andrey Kutepov}
\author[a,b]{Gabriel Kotliar}

\cortext[author] {Corresponding author.\\\textit{sachoi@bnl.gov} S.Choi}
\address[a]{Condensed Matter Physics and Materials Science Department,
Brookhaven National Laboratory, Upton, NY 11973, USA}
\address[b]{Department of Physics and Astronomy, Rutgers University, NJ 08854, USA}

\begin{abstract}

  ComDMFT is a massively parallel computational package to study the electronic structure of correlated-electron systems (CES). Our approach is a parameter-free method based on \textit{ab initio} linearized quasiparticle self-consistent GW (LQSGW) and dynamical mean field theory (DMFT). The non-local part of the electronic self-energy is treated within \textit{ab initio} LQSGW and the local strong correlation is treated within DMFT. In addition to \textit{ab initio} LQSGW+DMFT, charge self-consistent LDA+DMFT methodology is also implemented, enabling multiple methods in one open-source platform for the electronic structure of CES. This package can be extended for future developments to implement other methodologies to treat CES.

\end{abstract}


\end{frontmatter}




{\bf PROGRAM SUMMARY/NEW VERSION PROGRAM SUMMARY}

\begin{small}
\noindent
{\em Program Title: ComDMFT}                                          \\
{\em Licensing provisions(please choose one): GPLv3 }                                   \\
{\em Programming language: fortran90, C++, and Python}                                   \\


{\em Nature of problem:}\\
There is no open-source code based on \textit{ab initio} GW+EDMFT and related methodologies to support their theoretical advancement for the electronic structure of correlated electron systems.\\

{\em Solution method:}\\
We implemented \textit{ab initio} LQSGW+DMFT methodology, as a simplification of \textit{ab initio} GW+EDMFT, for the electronic structure of correlated electron systems. In addition, charge self-consistent LDA+DMFT methodology is also implemented, enabling the comparison of multiple methods for the electronic structure of correlated electron systems in one platform. \\

{\em Additional comments:}\\
ComDMFT is built on top of Wannier90 \cite{mostofi_Wannier90ToolObtaining_2008} and FlapwMBPT \cite{kutepov_LinearizedSelfconsistentQuasiparticle_2017} codes. \\
\\

\end{small}

\section{Introduction}
One of the most challenging but intriguing questions in the field of materials science is how to understand and calculate the physical properties of materials, starting from \textit{first principles}. This is a particularly difficult question in the context of systems with partially-filled $d\mhyphen$ and $f\mhyphen$ subshells, so-called correlated-electron systems (CES). Typically, CES shows large enhancements of the effective electron mass \cite{stewart_NonFermiliquidBehaviorElectron_2001} relative to Kohn-Sham bands structure calculated in the local density approximation (LDA) and related physical quantities such as the Sommerfeld coefficient and the coefficient of the quadratic term in the temperature dependence of the resistivity \cite{rice_ElectronElectronScatteringTransition_1968, kadowaki_UniversalRelationshipResistivity_1986, jacko_UnifiedExplanationKadowaki_2009}. For reviews of CES, please see these references \cite{georges_DynamicalMeanfieldTheory_1996,hewson_KondoProblemHeavy_1997,kotliar_StronglyCorrelatedMaterials_2004,kotliar_ElectronicStructureCalculations_2006,imada_ElectronicStructureCalculation_2010,anisimov_ElectronicStructureStrongly_2010,georges_StrongCorrelationsHund_2013}. 
 Microscopically, electrons in open $d\mhyphen$ and $f\mhyphen$ subshells tend to be localized on a short timescale but itinerant on a long timescale. This coexistence of localized and itinerant characters results in competition among different forms of long-range order, making CES extremely sensitive to small changes in their control parameters resulting in large responses. This makes theoretical understanding of these materials challenging, though it opens up exciting potential applications such as oxide electronics, high-temperature superconductors, and spintronic devices.

 Conventional band-picture-based \textit{ab initio} approaches are not particularly useful in the CES context. There are many success stories in the prediction, based on density functional theory (DFT) \cite{hohenberg_InhomogeneousElectronGas_1964,kohn_SelfConsistentEquationsIncluding_1965} and \textit{ab initio} quasiparticle GW \cite{hedin_NewMethodCalculating_1965,hybertsen_ElectronCorrelationSemiconductors_1986}, of the ground and excited state properties of normal metals and semiconductors. These successes are supported by software package developments since the 1990s including Quantum espresso\cite{giannozzi_QUANTUMESPRESSOModular_2009}, Paratec\cite{pfrommer_UnconstrainedEnergyFunctionals_1999}, Abinit\cite{gonze_RecentDevelopmentsABINIT_2016}, BerkeleyGW\cite{deslippe_BerkeleyGWMassivelyParallel_2012}, Vasp\cite{kresse_EfficiencyAbinitioTotal_1996a}, Siesta\cite{_SIESTAMethodInitio_}, OpenMX\cite{_OpenMXWebsite_}, Ecalj\cite{kotani_QuasiparticleSelfconsistentGW_2018}, Questaal\cite{_QuestaalHome_a}, Elk\cite{_ElkCode_}, Wien2k\cite{wien2k}, Fleur\cite{_JuelichFLEURProject_}, RSPt\cite{wills_FullPotentialElectronicStructure_2010}, and FlapwMBPT \cite{kutepov_LinearizedSelfconsistentQuasiparticle_2017}. However, such treatments do not sufficiently capture strongly correlated phenomena. In the band picture on the conceptual basis of Landau's Fermi-liquid theory \cite{pines_TheoryQuantumLiquids_1989}, the single-particle spectrum can be approximated by bands composed of itinerant dressed electrons as there is a one-to-one correspondence between bare particles and dressed particles with renormalized mass and finite lifetime. By contrast, electrons in open $d\mhyphen$ and $f\mhyphen$ subshells are neither localized nor itinerant. In addition to the low-energy coherent (itinerant) excitation spectra, the high-energy spectra of CES are dominated by incoherent (localized) excitation. The spectral weights associated with incoherent excitations (Hubbard bands) are not negligible and sometimes overwhelming as demonstrated by paramagnetic Mott insulators \cite{mott_DiscussionPaperBoer_1937}. 

 In an effort to understand the various interesting physics in CES, numerous theoretical frameworks have been pursued. Among these approaches, one of the most successful approaches is the dynamical mean-field theory (DMFT)\cite{metzner_CorrelatedLatticeFermions_1989,muller-hartmann_CorrelatedFermionsLattice_1989,brandt_ThermodynamicsCorrelationFunctions_1989a,janis_NewConstructionThermodynamic_1991,georges_HubbardModelInfinite_1992,jarrell_HubbardModelInfinite_1992,rozenberg_MottHubbardTransitionInfinite_1992,georges_NumericalSolutionEnsuremath_1992,georges_DynamicalMeanfieldTheory_1996}. DMFT establishes a mapping from a quantum many-body problem on a lattice onto a multi-orbital quantum impurity problem in an effective electron bath. Calculated objects associated with the impurity ($d\mhyphen$ and $f\mhyphen$ subshell) orbitals are embedded into the lattice to obtain single-particle Green’s function as well as susceptibilities. In combination with density functional theory \cite{anisimov_FirstprinciplesCalculationsElectronic_1997a,lichtenstein_InitioCalculationsQuasiparticle_1998, lichtenstein_FiniteTemperatureMagnetismTransition_2001,kotliar_ElectronicStructureCalculations_2006}, it has described various phenomena peculiar to CES such as paramagnetic Mott transitions\cite{held_MottHubbardMetalInsulatorTransition_2001} and volume collapse transitions\cite{savrasov_CorrelatedElectronsDplutonium_2001}. Now, there are many recent achievements within DFT+DMFT including free energy \cite{haule_FreeEnergyStationary_2015} and forces\cite{haule_ForcesStructuralOptimizations_2016}. For the double-counting within DFT+DMFT, a new scheme, so-called exact double-counting\cite{haule_ExactDoublecountingCombining_2015}, has been recently proposed.

All these theoretical advancements and their successes are supported by very recent software developments since the 2010s in various directions. Reusable physics libraries including TRIQS\cite{parcollet_TRIQSToolboxResearch_2015} and ALPS \cite{gaenko_UpdatedCoreLibraries_2017} and standalone impurity solvers such as iQIST\cite{huang_IQISTOpenSource_2015} and W2dynamics \cite{wallerberger_W2dynamicsLocalOne_2018} have been developed. Several DFT+DMFT packages have been developed: i) EDMFTF\cite{haule_DynamicalMeanfieldTheory_2010,_EmbeddedDMFTFunctional_} integrated with Wien2K, ii) TRIQS/DFTTools \cite{aichhorn_TRIQSDFTToolsTRIQS_2016} built on top of TRIQS and Wien2k, iii) VASP+DMFT \cite{park_ComputingTotalEnergies_2014} integrated with VASP \cite{kresse_EfficiencyAbinitioTotal_1996a} and EDMFTF package impurity solver \cite{haule_QuantumMonteCarlo_2007,haule_DynamicalMeanfieldTheory_2010,_EmbeddedDMFTFunctional_} iv) D-Core \cite{_DCoreDCoreDocumentation_} built on top of TRIQS, ALPS, Quantum Espresso \cite{giannozzi_QUANTUMESPRESSOModular_2009} and OpenMX\cite{_OpenMXWebsite_,ozaki_VariationallyOptimizedAtomic_2003}, v) AMULET \cite{_Amulet_} integrated with Quantum Espresso \cite{giannozzi_QUANTUMESPRESSOModular_2009} and Elk \cite{_ElkCode_} vi) Wien2k+W2dynamics. DFT+DMFT functionality has been added to existing \textit{ab initio} codes: Abinit\cite{gonze_RecentDevelopmentsABINIT_2016} and RSPt\cite{wills_FullPotentialElectronicStructure_2010,granas_ChargeSelfconsistentDynamical_2012}.

The theoretical success of DFT+DMFT in various CES has spiked the interest in realizing a diagrammatically controlled \textit{ab initio} approach to the quantum many-body problem in solids. The advantage of this approach is that the non-local part of electronic self-energy can be restored by adding the first non-trivial non-local diagrammatic correction to DMFT self-energy. In addition, the value of the parameters (Coulomb interaction tensor and double counting energy) can be determined accordingly, leaving only the choice the correlated orbitals to be made. 

There are now intense activities in this area with many recent studies on how to combine GW with extended DMFT (EDMFT) in a simplified way \cite{tomczak_CombinedGWDynamical_2012,sakuma_ElectronicStructureSrVO3_2013,taranto_ComparingQuasiparticleGW_2013,tomczak_AsymmetryBandWidening_2014,werner_DynamicalScreeningCorrelated_2016a,tomczak_QSGWDMFTElectronic_2015,roekeghem_ScreenedExchangeDynamical_2014,nilsson_MultitierSelfconsistentGW_2017,choi_FirstprinciplesTreatmentMott_2016,tomczak_MergingGWDMFT_2017}. However, in contrast to the DFT+DMFT case, there are only a few closed packages of any simplified form of GW+EDMFT to support its theoretical advancement: Questaal \cite{_QuestaalHome_a} supported by EDMFTF package impurity solver (without dynamical screening effect) \cite{haule_QuantumMonteCarlo_2007,haule_DynamicalMeanfieldTheory_2010,_EmbeddedDMFTFunctional_} and VASP\cite{kresse_EfficiencyAbinitioTotal_1996a,tomczak_MergingGWDMFT_2017} supported by W2dynamics \cite{wallerberger_W2dynamicsLocalOne_2018}. 

Recently, our group proposed an efficient way of combining \textit{ab initio} linearized quasiparticle self-consistent GW (LQSGW) and DMFT \cite{choi_FirstprinciplesTreatmentMott_2016}: we perform \textit{ab initio} LQSGW calculation \cite{kutepov_ElectronicStructurePu_2012,kutepov_GroundstatePropertiesSimple_2009} and then correct the local part of GW self-energy using DMFT. For the impurity orbital in the DMFT step, we choose a very localized orbital spanning a large energy window, which contains most strongly hybridized bands along with upper and lower Hubbard bands. Having chosen the shape of the correlated orbitals, all the other parameters are determined accordingly: double-counting energy within local GW approximation and Coulomb interaction tensor within constrained random phase approximation (cRPA)\cite{aryasetiawan_FrequencydependentLocalInteractions_2004}. This effort extends the open-source diagrammatic platform FlapwMBPT \cite{kutepov_LinearizedSelfconsistentQuasiparticle_2017} with DMFT approaches. This method enabled the \textit{first-principles} study of Mott insulators in both their paramagnetic and antiferromagnetic phases \cite{choi_FirstprinciplesTreatmentMott_2016} as well as a narrow-gap correlated semiconductor \cite{chikina_AnisotropicCorrelatedElectronic_2018} without any adjustable parameters. 

ComDMFT is the first open-source massively parallel computer package that implements any simplified form of GW+EDMFT. Among them, our choice is LQSGW+DMFT methodology. In addition, ComDMFT enables multiple methods for the electronic structure of CES in one platform. Charge self-consistent LDA+DMFT methodology is implemented with the same linearized-augmented planewave (LAPW) basis set under the full potential from nuclei and the same way of choosing correlated orbitals from Wannier functions. It will be extended for rotationally-invariant-slave-boson methods (RISB) in combination with LDA. This package is built on top of FlapwMBPT code \cite{kutepov_LinearizedSelfconsistentQuasiparticle_2017} for \textit{ab initio} LQSGW and LDA calculations and composed of several programs: ComWann, ComCoulomb, ComDC, ComLowH, and ComCTQMC which are written in FORTRAN90, C++, and python. The algorithms and codes were devised and developed at Rutgers University, University of Sherbrooke, and Brookhaven National Lab. The source codes and online tutorials as a part of COMSUITE can be found at https://www.bnl.gov/comscope/software/comsuite.php.

The outline of the paper is as follows. We will introduce a full GW+EDMFT methodology in Sec. \ref{sec_gen_theo_framework}, and \textit{ab initio} LQSGW+DMFT methodology as a simplification of full GW+EDMFT in Sec. \ref{sec_lqsgw_dmft}. In Sec. \ref{sec_com_layout} and \ref{sec_lda_dmft},  we will describe computational layout for two different methodology available in ComDMFT: \textit{ab initio} LQSGW+DMFT implementation and charge self-consistent LDA+DMFT implementation. Benchmark results on NiO, MnO, and FeSe will be provided in Sec. \ref{sec_results}

\begin{table}
\caption{\label{tab:symbol}list of symbols}
\begin{tabular}{ll}
  $ \mathbf{k}$ & crystal momentum vector\\
  $ \mathbf{R}$ & lattice vector \\
  $ \omega_n$ & fermionic Matsubara frequency \\
  $ \nu_n$ & bosonic Matsubara frequency \\
  $ \tau$ & imaginary time \\
  $ N_\mathbf{k}$ & the number of crystal momentum vectors in the first Brillouin zone\\
  $ G$ & Green's function\\
  $ \Sigma$ & electronic self-energy\\
  $ W$ & screened Coulomb interaction\\
  $ W_r$ & partially screened Coulomb interaction\\      
  $ P$ & polarizability\\

  $ \chi$ & susceptibility\\
  $ f_\mathbf{k}$ & fermionic projection operator\\
  $ b_\mathbf{k}$ & bosonic projection operator\\  
  $ \widetilde{\mathcal{G}}$ & fermionic Weiss field\\
  $ \widetilde{G}_{imp}$ & impurity Green's function\\
  $ \widetilde{G}_{loc}$ & local Green's function\\     
  $ \widetilde{\Sigma}_{imp}$ & impurity self-energy\\
  $ \widetilde{\Sigma}_{DC}$ & double-counted self-energy\\
  $ \widetilde{\mathcal{U}}$ & bosonic Weiss field\\
  $ \widetilde{W}_{imp}$ & impurity screened Coulomb interaction\\
  $ \widetilde{W}_{loc}$ & local screened Coulomb interaction\\           
  $ \widetilde{P}_{imp}$ & impurity polarizability\\
  $ \widetilde{P}_{DC}$ & double-counted polarizability\\  
  $ \widetilde{\chi}_{imp}$ & impurity susceptibility\\

  $ G_{MF}$ & mean-field Green's function\\
  $ H_{MF}$ & mean-field Hamiltonian\\          
  $ G_{QP}$ & Green's function from a quasiparticle Hamiltonian\\
  $ P_{QP}$ & polarizability from a quasiparticle Green's function\\
  $ P_{QP}^{low}$ & polarizability from correlated bands\\
  $ P_{QP}^{high}$ & $ P_{QP}- P_{QP}^{low}$\\
  $N,M$ & quasiparticle band index\\

  $ E_{N\mathbf{k}}$ & quasiparticle energy with a band index $N$ and a crystal momentum vector $\mathbf{k}$\\  
  $| \psi_{N\mathbf{k}}\rangle$ & quasiparticle wave function ($\psi_{N\mathbf{k}}(\mathbf{r})=\langle\mathbf{r}|\psi_{N\mathbf{k}} \rangle$) \\
  $I,J,K,L$ & Wannier orbital composite index for the centered atom\\
                & and approximate angular momentum quantum numbers of the orbital. \\
                & If the index is a small letter, the orbital is a correlated one.\\    
  $| W_{I\mathbf{R}} \rangle$ & Wannier function labeled by lattice vector $\mathbf{R}$ and an orbital orbital $I$.\\
  $| W_{I\mathbf{k}} \rangle$ & Bloch sum of Wannier functions $| W_{I \mathbf{R}} \rangle$.\\
  $U_{NI}(\mathbf{k})$ & basis rotation matrix $\langle \psi_{N \mathbf{k}} | W_{I \mathbf{k}} \rangle$.\\
  $F^k$ & Slater's integral of partially screened Coulomb interaction\\    
\end{tabular}
\end{table}

\section{General Theoretical Framework} \label{sec_gen_theo_framework}

\begin{figure}
  \centering
  \includegraphics[width=1.0\columnwidth]{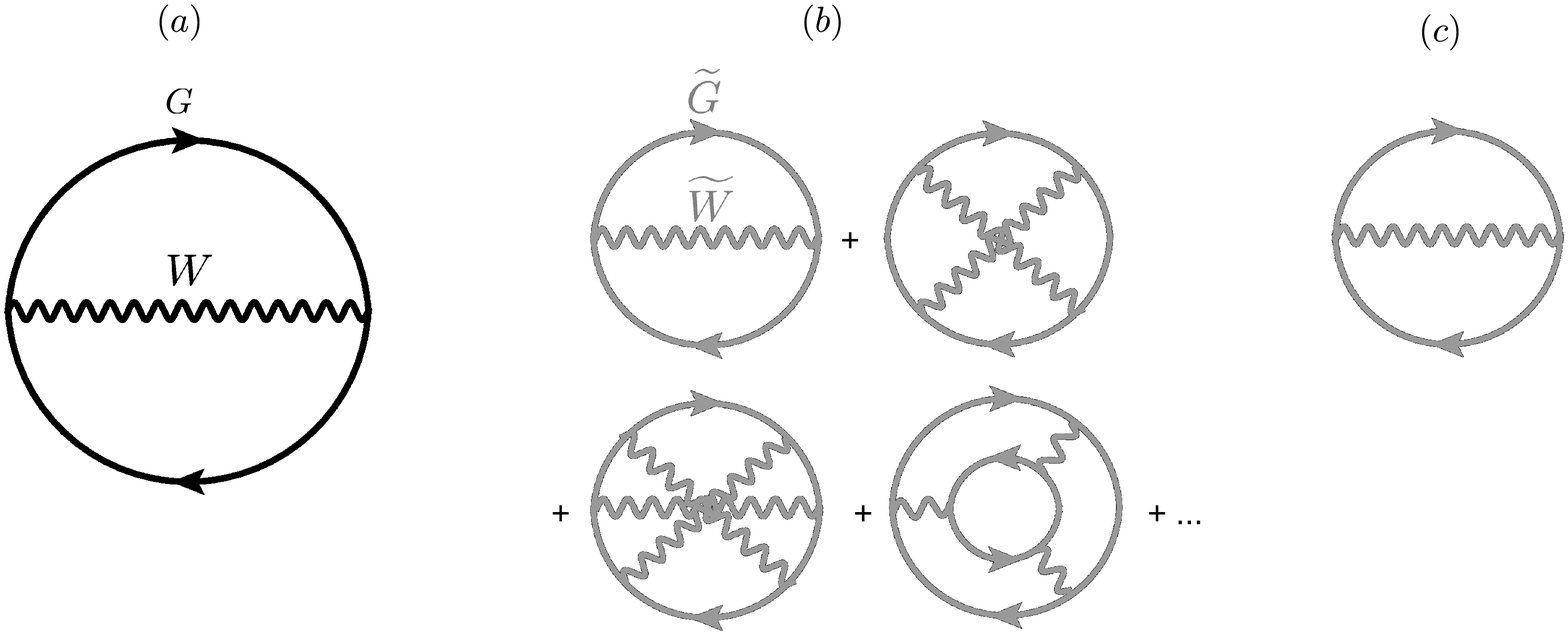}
  \caption{Diagrammatic representations of three terms in full GW+EDMFT $\Psi$ functionals. (a) GW $\Psi$  functional (b) EDMFT $\Psi$ functional and (c) local-GW $\Psi$ functional. Full and wiggly lines represent Green's function ($G$) and screened Coulomb interaction ($W$), respectively. Black and gray colors represent quantities in the full Hilbert space and in the correlated subspace, respectively.}
  \label{fig_functional}
\end{figure}

After constructing a Baym-Kadanoff functional of the single-particle Green's function ($G$) and the screened Coulomb interaction ($W$)\cite{almbladh_VariationalTotalEnergies_1999,chitra_EffectiveactionApproachStrongly_2001}, various approximations to the so-called $\Psi$ functional can be made: the first order approximation in $W$ corresponds to the GW approximation ($\Psi^{GW}(G,W)= -\frac{1}{2}\Tr GWG$)\cite{almbladh_VariationalTotalEnergies_1999}, whereas the local approximation applied to the $d\mhyphen$ or $f\mhyphen$ subshell orbitals gives rise to EDMFT approach \cite{si_KosterlitzThoulessTransitionShort_1996,sengupta_NonFermiliquidBehaviorSpinglass_1995,henrikkajueter_InterpolatingPerturbationScheme_1996}($\Psi(\widetilde{G},\widetilde{W})$). By combining these two diagrammatic approaches \cite{sun_ExtendedDynamicalMeanfield_2002,biermann_FirstPrinciplesApproachElectronic_2003,nilsson_MultitierSelfconsistentGW_2017}, we can derive the $\Psi$ functional of GW+EDMFT ($\Psi^{GW+EDMFT}(G,W)$) after subtracting the double-counted local GW diagram ($\Psi^{local-GW}(\widetilde{G},\widetilde{W})= -\frac{1}{2}\Tr \widetilde{G}\widetilde{W}\widetilde{G}$), as shown in Fig. \ref{fig_functional}.
\begin{equation}
  \begin{split}
    \Psi^{GW+EDMFT}(G,W)= -\frac{1}{2}\Tr GWG+\Psi(\widetilde{G},\widetilde{W})+\frac{1}{2}\Tr \widetilde{G}\widetilde{W}\widetilde{G}.
  \end{split}
\end{equation}
Here $\widetilde{A}$ represents the projection of the quantity $A$ to the correlated subspace.

\begin{figure}
  \centering
  \includegraphics[width=1.0\columnwidth]{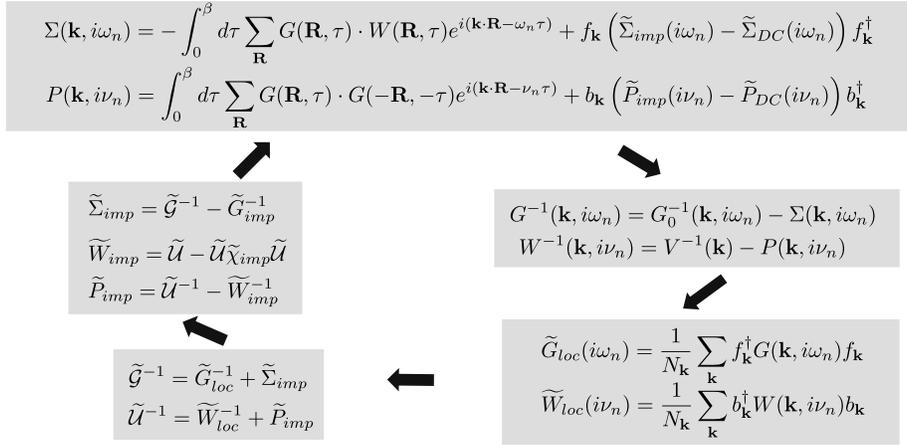}
  \caption{general theoretical framework for \textit{ab initio}  GW+EDMFT. For the meaning of each symbol, please see table \ref{tab:symbol}. Starting from GW self-energy and polarizability, impurity self-energy ($\widetilde{\Sigma}_{imp}$) and impurity polarizability ($\widetilde{P}_{imp}$) are embedded with proper double-counting terms ($\widetilde{\Sigma}_{DC}$ and $\widetilde{P}_{DC}$) by using fermionic ($f_\mathbf{k}$) and bosonic ($b_\mathbf{k}$) projection operators to the correlated subspace, respectively. Next, single-particle Greens function ($G$) and fully screened Coulomb interaction ($W$) are obtained and their projections to the correlated subspace ($\widetilde{G}_{loc}$ and $\widetilde{W}_{loc}$) are calculated. By using impurity self-energy ($\widetilde{\Sigma}_{imp}$) and impurity polarizability ($\widetilde{P}_{imp}$) from the previous iteration, fermionic ($\widetilde{\mathcal{G}}$) and bosonic ($\widetilde{\mathcal{U}}$) Weiss fields are calculated. With the inputs $\widetilde{\mathcal{G}}$ and $\widetilde{\mathcal{U}}$, impurity Greens function ($\widetilde{G}_{imp}$) and impurity susceptibility ($\widetilde{\chi}_{imp}$) are calculated and then new $\widetilde{\Sigma}_{imp}$ and $\widetilde{P}_{imp}$ are extracted. The self-consistent loop is continued until $\widetilde{G}_{imp}=\widetilde{G}_{loc}$ and $\widetilde{W}_{imp}=\widetilde{W}_{loc}$.}
  \label{fig_gw_edmft_scf}
\end{figure}

Within the fully self-consistent GW+EDMFT, the electronic structure can be calculated by the self-consistent loop in Fig. \ref{fig_gw_edmft_scf}. Starting from irreducible self-energy ($\Sigma$) and polarizability ($P$) within GW approximation, we embed impurity self-energy ($\widetilde{\Sigma}_{imp}$) and impurity polarizability ($\widetilde{P}_{imp}$) with proper double-counting terms ($\widetilde{\Sigma}_{DC}$ and $\widetilde{P}_{DC}$) by using fermionic ($f_\mathbf{k}$) and bosonic ($b_\mathbf{k}$) projection operators \cite{biermann_FirstPrinciplesApproachElectronic_2003} to the correlated subspace. If it is the first iteration, we may set impurity quantities the same as the double-counting quantities. Then, single-particle Greens function ($G$) and fully screened Coulomb interaction ($W$) can be obtained using Dyson's equation. After projecting $G$ and $W$ to the correlated subspace using $f_\mathbf{k}$ and $b_\mathbf{k}$, local Green's function ($\widetilde{G}_{loc}$) and local screened Coulomb interaction ($\widetilde{W}_{loc}$) are obtained. By using impurity self-energy ($\widetilde{\Sigma}_{imp}$) and impurity polarizability ($\widetilde{P}_{imp}$) from the previous iteration, fermionic ($\widetilde{\mathcal{G}}$) and bosonic ($\widetilde{\mathcal{U}}$) Weiss fields are calculated. With the inputs $\widetilde{\mathcal{G}}$ and $\widetilde{\mathcal{U}}$, impurity Greens function ($\widetilde{G}_{imp}$) and impurity susceptibility ($\widetilde{\chi}_{imp}$) are calculated and then new $\widetilde{\Sigma}_{imp}$ and $\widetilde{P}_{imp}$ are extracted. We solve this self-consistent equation until $\widetilde{G}_{imp}=\widetilde{G}_{loc}$ and $\widetilde{W}_{imp}=\widetilde{W}_{loc}$.

\section{\textit{Ab initio} LQSGW+DMFT as a simplified version of GW+EDMFT} \label{sec_lqsgw_dmft}

Although this is a well-known route, its implementation has been realized \cite{nilsson_MultitierSelfconsistentGW_2017} only in the low-energy Hilbert space for GW+EDMFT. Furthermore, as is often the case in quantum many-body theory, approximations which are not fully diagrammatic or not fully self-consistent often work better for some quantities than the fully diagrammatic treatments. For example, fully self-consistent GW is worse than partially self-consistent GW for the spectra of three-dimensional electron gas \cite{holm_FullySelfconsistentMathrmGW_1998}. In addition, for the Anderson impurity model, it has been shown that there is substantial cancellation of each term in the same expansion order with respect to onsite interaction and all the diagrams should be collected to leave a very small value for the expansion coefficient. \cite{yosida_PerturbationExpansionAnderson_1970,yamada_PerturbationExpansionAnderson_1975a,yosida_PerturbationExpansionAnderson_1975,yamada_PerturbationExpansionAnderson_1975}. So far various partial self-consistency schemes have been tried \cite{sun_ManyBodyApproximationScheme_2004,tomczak_CombinedGWDynamical_2012,sakuma_ElectronicStructureSrVO3_2013,taranto_ComparingQuasiparticleGW_2013,tomczak_AsymmetryBandWidening_2014,roekeghem_ScreenedExchangeDynamical_2014,tomczak_QSGWDMFTElectronic_2015,nilsson_MultitierSelfconsistentGW_2017,choi_FirstprinciplesTreatmentMott_2016}. In the most cases, GW+EDMFT self-consistency for the bosonic quantities has been neglected and the bosonic Weiss field is obtained from cRPA and its various extensions. In addition, Green's functions are assumed to be $G^{-1}(\mathbf{k},i\omega_n)=G_{MF}^{-1}(\mathbf{k},i\omega_n)-f_\mathbf{k}\widetilde{\Sigma}_{imp}(i\omega_n)f_\mathbf{k}^\dagger$, where $G_{MF}$ is a free Green's function from a mean-field Hamiltonian ($H_{MF}$) such that $G_{MF}^{-1}(\mathbf{k},i\omega_n)=i\omega_n-H_{MF}(\mathbf{k})$. $G_{MF}$ is fixed during the self-consistent loop implying that mean-field self-energy is corrected by one-shot DMFT.

\begin{figure}
  \centering
  \includegraphics[width=1.0\columnwidth]{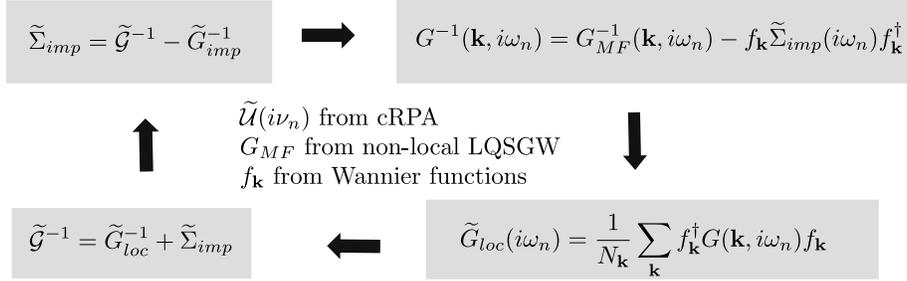}
  \caption{ \textit{ab initio} LQSGW+DMFT self-consistent loop, implemented in ComDMFT, as a simplified version of \textit{ab initio} GW+EDMFT depicted in Fig \ref{fig_gw_edmft_scf}. For the meaning of each symbol, please see table \ref{tab:symbol}. With a fixed mean-field Green's function ($G_{MF}$) within \textit{ab initio} non-local LQSGW, a fixed bosonic Weiss field ($\widetilde{\mathcal{U}}$) from a constrained random phase approximation (see Sec. \ref{sec_comcoulomb}), and a choice of correlated orbitals ($f_\mathbf{k}$, see Sec. \ref{sec_comwann}) , we solve dynamical mean-field theory self-consistent equation. Onto $G_{MF}$ in which double-counting self-energy is compensated as described in Sec. \ref{sec_comlowh}, we embed $\widetilde{\Sigma}_{imp}$ using $f_\mathbf{k}$. After projecting $G$ to the correlated subspace using $f_\mathbf{k}$, local Green's function ($\widetilde{G}_{loc}$) is obtained. By using impurity self-energy ($\widetilde{\Sigma}_{imp}$) from the previous iteration, $\widetilde{\mathcal{G}}$ is calculated. With the inputs of $\widetilde{\mathcal{G}}$ and $\widetilde{\mathcal{U}}$, $\widetilde{G}_{imp}$ is calculated and subsequently a new $\widetilde{\Sigma}_{imp}$ is obtained. We solve this self-consistent equation until $\widetilde{G}_{imp}=\widetilde{G}_{loc}$.
  }
  \label{fig_lqsgw_dmft_scf}
\end{figure}

For the construction of $H_{MF}$, various static mean-field approaches have been suggested and tested, ranging from one-shot GW \cite{taranto_ComparingQuasiparticleGW_2013,tomczak_AsymmetryBandWidening_2014}, screened exchange \cite{roekeghem_ScreenedExchangeDynamical_2014}, LQSGW\cite{choi_FirstprinciplesTreatmentMott_2016} and QSGW\cite{sponza_NonlocalSelfenergiesMetals_2016}. In addition, the idea of the non-local version of QSGW and LQSGW was proposed\cite{tomczak_QSGWDMFTElectronic_2015}. Among them, our choice of $H_{MF}$ is \textit{ab initio} non-local LQSGW. LQSGW is known to fix inaccurate spectra of fully self-consistent GW \cite{holm_FullySelfconsistentMathrmGW_1998} and, at the same time, the starting point dependence of one-shot GW \cite{chen_BandedgePositionsGW_2014}. By using non-local LQSGW Hamiltonian, the effect of double-counting self-energy is compensated up to linear order in frequency \cite{tomczak_QSGWDMFTElectronic_2015}.

Starting from $G_{MF}$ within \textit{ab initio} non-local LQSGW, the bosonic Weiss field from cRPA and the choice of correlated orbitals, we solve DMFT self-consistent equation as shown in Fig. \ref{fig_lqsgw_dmft_scf}. Onto $G_{MF}$, we embed $\widetilde{\Sigma}_{imp}$ using $f_\mathbf{k}$. If it is the first iteration, we may set impurity quantities the same as the double-counting quantities. After projecting $G$ to the correlated subspace using $f_k$, local Green's function ($\widetilde{G}_{loc}$) is obtained. By using impurity self-energy ($\widetilde{\Sigma}_{imp}$) from the previous iteration, $\widetilde{\mathcal{G}}$ is calculated. With the inputs of $\widetilde{\mathcal{G}}$ and $\widetilde{\mathcal{U}}$, $\widetilde{G}_{imp}$ is calculated and subsequently a new $\widetilde{\Sigma}_{imp}$ is obtained. We solve this self-consistent equation until $\widetilde{G}_{imp}=\widetilde{G}_{loc}$. 

\section{Computational Layout for \text{ab initio} LQSGW+DMFT}\label{sec_com_layout}
\begin{figure}
  \centering
  \includegraphics[width=1.0\columnwidth]{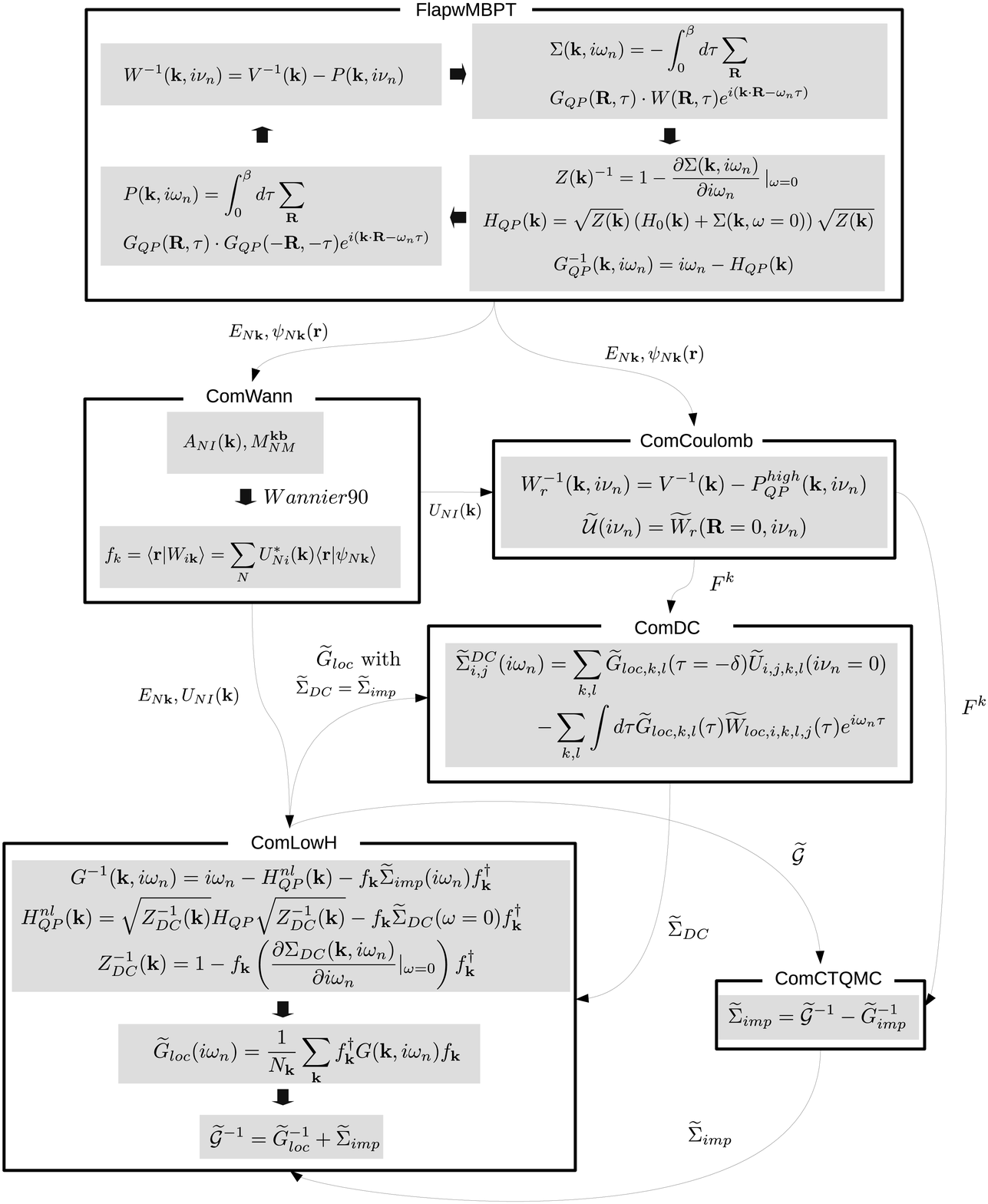}
  \caption{ \textit{ab initio} LQSGW+DMFT flowchart and data exchange between programs (FlapwMBPT, ComWann, ComCoulomb, ComDC, ComLowH and ComCTQMC) in ComDMFT. For the meaning of each symbol, please see table \ref{tab:symbol}. First, a quasiparticle Hamiltonian ($H_{QP}$) is constructed within \textit{ab initio} linearized quasiparticle self-consistent GW (LQSGW) approximation by using FlapwMBPT \cite{kutepov_LinearizedSelfconsistentQuasiparticle_2017}. Within LQSGW, GW self-energy and polarizability within random phase approximation are constructed by using LQSGW Green's function ($G_{QP}$). With LQSGW quasiparticle energies and wavefunctions, ComWann calculates initial trial rotation matrix ($A_{NI}(\mathbf{k})$) and overlap matrix ($M_{NM}^{\mathbf{k}\mathbf{b}}$) to construct Wannier functions and their rotation matrix ($U_{NI}(\mathbf{k})$) by using Wannier90 package \cite{mostofi_Wannier90ToolObtaining_2008}. ComCoulomb reads LQSGW bandstructure as well as $U_{NI}(\mathbf{k})$ to calculate Slater's integrals ($F^{k}$) of partially screened Coulomb interaction ($W_r$) within a constrained random phase approximation. Then ComDC calculates self-energy within local GW approximation with inputs of $F^{k}$ and LQSGW local Green's function ($\widetilde{G}_{loc}$). Finally, we solve DMFT self-consistent loop by using ComLowH and ComCTQMC. ComLowH Wannier-interpolates LQSGW bandstructure and construct Lattice Green's function ($G$) and fermionic Weiss field ($\widetilde{\mathcal{G}}$) by compensating double-counting self-energy up to linear order in frequency and embedding impurity self-energy ($\widetilde{\Sigma}_{imp}$). ComCTQMC calculates impurity self-energy with inputs of $\widetilde{\mathcal{G}}$ and $F^{k}$}
  \label{fig_flow}
\end{figure}


The \textit{ab initio} LQSGW+DMFT approach is an approximation of \textit{ab initio} GW+EDMFT. Input parameters are the crystal structure and the chosen correlated orbitals. The LQSGW+DMFT flowchart is provided in Fig. \ref{fig_flow}. \textit{Ab initio} LQSGW+DMFT calculations can be broken down into five steps: i) the construction of quasi-particle Hamiltonian within \textit{ab initio} LQSGW by FlapwMBPT \cite{kutepov_LinearizedSelfconsistentQuasiparticle_2017} corresponding to Fig. \ref{fig_functional} (a) ii) the construction of the atom-centered local basis set spanning the low-energy Hilbert space by ComWann utilizing Wannier90 package \cite{mostofi_Wannier90ToolObtaining_2008} iii) the calculation of the bosonic Weiss field within cRPA and evaluation of Slater's integral associated with the impurity orbitals by ComCoulomb iv) the calculation of the double-counted self-energy associated with the impurity orbitals within local GW approximation by ComDC corresponding to Fig. \ref{fig_functional} (c), v) Wannier interpolation of the mean-field Hamiltonian and solving the DMFT self-consistent equation by ComLowH and ComCTQMC corresponding to Fig. \ref{fig_functional} (b). 

\subsection{ComWann}\label{sec_comwann}
In the one-particle picture, the electronic structure of a periodic system can be represented by extended states. Among the possible choices is quasiparticle wavefunctions ($|\psi_{N\mathbf{k}}\rangle$), which are eigenvectors of an effective one-particle Hamiltonian. $|\psi_{N\mathbf{k}}\rangle$ are characterized by two good quantum numbers: band index $N$ and crystal momentum vector $\mathbf{k}$. Alternative basis sets composed of localized states are also available and this choice is useful in representing the low-energy Hilbert space. One of the typical choices of localized states is Wannier functions ($|W_{I\mathbf{R}}\rangle$)\cite{marzari_MaximallyLocalizedWannier_2012}, as labeled by a lattice vector $\mathbf{R}$ and a composite index $I$ for the centered atom and approximate angular momentum quantum numbers of the wave function. $|W_{I\mathbf{R}}\rangle$ is constructed in the following way. 
\begin{equation}
  \begin{split}
    |W_{I\mathbf{R}}\rangle=\frac{1}{\sqrt{N_\mathbf{k}}}\sum_{\mathbf{k},N}U_{NI}^*(\mathbf{k})e^{-i\mathbf{k}\cdot\mathbf{R}}|\psi_{N\mathbf{k}}\rangle,
    \label{eq:wannier_fun}    
  \end{split}    
\end{equation}
where $N_\mathbf{k}$ is the number of crystal momentum vector in the first Brillouin zone. We also use $|W_{I\mathbf{k}}\rangle=\sum_{N}U_{NI}^*(\mathbf{k})|\psi_{N\mathbf{k}}\rangle$ in the following. One way to construct orthonormal basis set $|W_{I\mathbf{k}}\rangle$ from $|\psi_{N\mathbf{k}}\rangle$, or to determine $U_{NI}(\mathbf{k})$, is by minimizing total spreads ($\Omega$) defined by
\begin{equation}
  \begin{split}
    \Omega=\sum_{I \mathbf{R}}\langle\mathbf{r}^2-\langle\mathbf{r}\rangle_{I\mathbf{R}}^2\rangle_{I\mathbf{R}},
  \end{split}    
\end{equation}
under the constraint that eigenvalues $E_{N\mathbf{k}}$ in an energy window (so-called frozen energy window) is preserved. Here $\langle A\rangle_{I\mathbf{R}}=\langle W_{I\mathbf{R}}|A|W_{I\mathbf{R}}\rangle$ and $\mathbf{r}$ is the position vector in a global coordinate system.

In ComDMFT,  $|W_{I\mathbf{R}}\rangle$ is constructed in the following way. i) ComWann sets the initial $U_{NI}(\mathbf{k})$ or pick out initial trial orbitals $|T_{I\mathbf{R}}\rangle$ for $\Omega$ minimization. A radial function of an initial trial orbital with an angular momentum character can be any linear combination of muffin-tin orbitals in the LAPW basis set with the angular momentum character. ComWann takes the linear combination which maximize  $\sum_{N,\mathbf{k}}^{E_{froz}^{min}<E_{N,\mathbf{k}}<E_{froz}^{max}} |\langle \psi_{N\mathbf{k}}|T_{I\mathbf{k}}\rangle|^2$, where $E_{froz}^{min}$ is the lower bound of the frozen energy window and $E_{froz}^{max}$ is the upper bound of the frozen energy window. Among these muffin-tin orbitals with the choice of radial function, ComWann picks out one of which $\frac{1}{N_\mathbf{k}}\sum_{N,\mathbf{k}}^{E_{froz}^{min}<E_{N\mathbf{k}}<E_{froz}^{max}} |\langle \psi_{N\mathbf{k}}|T_{I\mathbf{k}}\rangle|^2> 0.15 $ and set them initial trial orbitals. ii) Then ComWann calculates the overlap matrix $M_{NM}^{\mathbf{k}\mathbf{b}}=\langle u_{N\mathbf{k}} | u_{M\mathbf{k}+\mathbf{b}}\rangle$. Here $|u_{N\mathbf{k}}\rangle$ is the periodic part of the $|\psi_{N\mathbf{k}}\rangle$ and $\mathbf{b}$ is a vector connecting a $\mathbf{k}$ point to its nearest neighbors in the $\mathbf{k}$ space \cite{marzari_MaximallyLocalizedWannier_2012}. $M_{NM}^{\mathbf{k}\mathbf{b}}$ enables the calculation of the total spread in the crystal momentum space \cite{marzari_MaximallyLocalizedWannier_2012}. iii) $\Omega$ minimization is carried out by using Wannier90 package in its library mode \cite{mostofi_Wannier90ToolObtaining_2008} by providing so-called A matrix $A_{NI}(\mathbf{k})=\frac{1}{N_\mathbf{k}}\sum_{\mathbf{k}}\langle \psi_{N\mathbf{k}}|T_{I\mathbf{k}}\rangle$ and $M_{NM}^{\mathbf{k}\mathbf{b}}$. iv) Once $U_{NI}(\mathbf{k})$ is determined, $f_\mathbf{k}$  is given by
\begin{equation}
  \begin{split}
f_\mathbf{k}=\langle\mathbf{r}|W_{i\mathbf{k}}\rangle=\sum_{N}U_{Ni}^*(\mathbf{k})\langle\mathbf{r}|\psi_{N\mathbf{k}}\rangle
  \end{split}    
\end{equation}
where a small letter $i$ for the wannier orbital indicates the wannier orbitals is for a correlated orbital.

In ComWann, the default choice of the approximate angular momentum character of Wannier functions depends on the presence of spin-orbit coupling. For a system without spin-orbit coupling, the angular part is approximately 
the cubic-harmonics $Y_{lm}$, which is defined as
\begin{equation}
  \begin{split}
    Y_{lm}=\begin{cases}
      \frac{i}{\sqrt{2}} \left(Y_{l}^{-|m|}-(-1)^mY_{l}^{|m|}\right),& m<0\\
      Y_{l}^0,& m=0\\
      \frac{1}{\sqrt{2}} \left(Y_{l}^{-|m|}+(-1)^mY_{l}^{|m|}\right),& m>0\\      
      \end{cases}
    \label{eq:s_wso}
  \end{split}
\end{equation}
Here, $Y_{l}^{m}$ is the spherical harmonics. Note that subscript $m$ denotes cubic harmonics and superscript $m$ denotes spherical harmonics. For the system with spin-orbit coupling, the angular part is approximately 
the spin-angular function ($\Omega_{l,a,m}$) which is defined as
\begin{equation}
  \begin{split}
    \Omega_{l,a=\pm\frac{1}{2},m}=\sum_{s\pm 1/2}C_{a,s}^{l,m}Y_{l}^{m-s}(\hat{r})u_s.
  \end{split}
\end{equation}
Here, $u_s$ is a spinor and $C_{a,s}^{l,m}=\langle l, m-s,\frac{1}{2}, s | l+a, m \rangle$.

Parallelization of ComWann is characterized by the calculation of $A_{NI}(\mathbf{k})$, the calculation of $M_{NM}^{\mathbf{k}\mathbf{b}}$ and total spread minimization.  For the computation of $A_{NI}(\mathbf{k})$ and $M_{NM}^{\mathbf{k}\mathbf{b}}$, the code is MPI-parallelized over $\mathbf{k}$ with with nearly linear scaling up to $N_\mathbf{k}$ processors. 
For the minimization of the total spread, ComWann calls Wannier90 package in its library mode \cite{mostofi_Wannier90ToolObtaining_2008}, which utilize only a single processor.

\subsection{ComCoulomb}\label{sec_comcoulomb}

ComCoulomb calculates the bosonic Weiss field $\widetilde{\mathcal{U}}$ associated with the correlated orbitals within cRPA\cite{aryasetiawan_FrequencydependentLocalInteractions_2004,aryasetiawan_CalculationsHubbardFirstprinciples_2006} and its Slater's integrals. Here we stress that the bosonic Weiss field $\widetilde{\mathcal{U}}$ from cRPA is a way to evaluate $\widetilde{\mathcal{U}}$ and not identical to $\widetilde{\mathcal{U}}$ from “ideal” fully self-consistent GW+EDMFT defined in Fig. \ref{fig_gw_edmft_scf}.  Within cRPA, the bosonic Weiss field $\widetilde{\mathcal{U}}$ is obtained by separating out the RPA polarizability ($P_{QP}^{low}$) from the correlated states. Various ideas to identify  $P_{QP}^{low}$ have been suggested ranging from the usage of energy window \cite{aryasetiawan_CalculationsHubbardFirstprinciples_2006,miyake_ScreenedCoulombInteraction_2008}, the disentanglement of correlated bands from itinerant bands \cite{miyake_InitioProcedureConstructing_2009}, weighting the polarizability by correlated orbital projections \cite{sasioglu_EffectiveCoulombInteraction_2011,shih_ScreenedCoulombInteraction_2012}, selecting correlated bands \cite{aryasetiawan_CalculationsHubbardFirstprinciples_2006,miyake_ScreenedCoulombInteraction_2008,werner_SatellitesLargeDoping_2012}, to local polarizability \cite{nomura_EffectiveOnsiteInteraction_2012}. In ComCoulomb, $P_{QP}^{low}$ is defined by identifying correlated bands \cite{aryasetiawan_CalculationsHubbardFirstprinciples_2006,miyake_ScreenedCoulombInteraction_2008,werner_SatellitesLargeDoping_2012}: bands having strong correlated orbitals character are chosen as correlated bands at each $\mathbf{k}$ point and the number of correlated bands is set to be the number of correlated orbitals. Then, $P_{QP}^{low}$ is defined in the following way. 

\begin{equation}
  \begin{split}
    P_{QP}^{low}(\mathbf{r},\mathbf{r}',\mathbf{k},i\nu_n)=-N_s\sum_{\mathbf{k'}}\sum_{N}^{\substack{\text{unoccupied }\\\text{correlated bands}}}\sum_{M}^{\substack{\text{occupied}\\\text{correlated bands}}}\\
    \psi_{N\mathbf{k}'}(\mathbf{r})\psi_{M\mathbf{k}'+\mathbf{k}}^*(\mathbf{r})\psi_{N\mathbf{k}'}^*(\mathbf{r}')\psi_{M\mathbf{k}'+\mathbf{k}}(\mathbf{r}')\frac{2(E_{N\mathbf{k}'}-E_{M\mathbf{k}'+\mathbf{k}})}{\nu_n^2-(E_{N\mathbf{k}'}-E_{M\mathbf{k}'+\mathbf{k}})^2},
    \label{eq:chi_low}
  \end{split}
\end{equation}
where $\psi_{N\mathbf{k}}(\mathbf{r})=\langle \mathbf{r}|\psi_{N\mathbf{k}}\rangle$ and $N_s$ is 2 for the system without spin-orbit coupling and 1 for the system with strong spin-orbit coupling. Using $P_{QP}^{high}=P_{QP}-P_{QP}^{low}$ where $P_{QP}$ is RPA polarizability, the partially-screened Coulomb interaction ($W_r$) is calculated by
\begin{equation}
  \begin{split}
    W_r^{-1}(\mathbf{r},\mathbf{r}',\mathbf{k},i\nu_n)=V^{-1}(\mathbf{r},\mathbf{r}',\mathbf{k})-P_{QP}^{high}(\mathbf{r},\mathbf{r}',\mathbf{k},i\nu_n).
    \label{eq:partial_coulomb}    
  \end{split}
\end{equation}
Next, Slater's integrals ($F^k$) \cite{slater_QuantumTheoryAtomic_1960,sugano_MultipletsTransitionMetalIons_2012} are calculated using $W_r(\mathbf{r},\mathbf{r}',\mathbf{R}=0,i\nu_n)$ and Wannier functions for the correlated orbitals. Here we note that Slater parameterization of Coulomb interaction tensor is an approximation by assuming full rotation symmetry. Another parameterization including crystal field splitting effect is also possible \cite{ribic_CubicInteractionParameters_2014}. For the system without spin-orbit coupling, 
\begin{equation}
  \begin{split}
    F^k(i\nu_n)=&\frac{1}{C_{k}^l}\frac{4\pi}{2k+1}\sum_{m_1,m_2,m_3,m_4}\sum_{m'_1,m'_2,m'_3,m'_4}\\
    &\langle Y_l^{m_1}|Y_k^{m_1-m_4}Y_l^{m_4}\rangle\langle Y_l^{m_2}Y_k^{m_2-m_3}|Y_l^{m_3}\rangle\\
    &S_{m_1m'_1}S_{m_2m'_2}S^{-1}_{m_3m'_3}S^{-1}_{m_4m'_4}
    \int d\mathbf{r}d\mathbf{r}'W_r(\mathbf{r},\mathbf{r}', R=0,i\nu_n)\\
    &W_{R=0,m'_1}^*(\mathbf{r})W_{R=0,m'_2}^*(\mathbf{r}')W_{R=0,m'_3}(\mathbf{r}')W_{R=0,m'_4}(\mathbf{r}),
    \label{eq:slater_woso}        
  \end{split}
\end{equation}
where $C_k^l=\frac{(2l+1)^4}{{2k+1}}\begin{bmatrix} l & k & l \\ 0 & 0 & 0\end{bmatrix}$, $\begin{bmatrix} l & k & l \\ 0 & 0 & 0\end{bmatrix}$ is the Racah-Wigner 3j-symbol, $S_{m_1m'_1}=\langle Y_{lm_1}|Y_{l}^{m'_1}\rangle$ and $W_{\mathbf{R},m}(\mathbf{r})$ is a correlated Wannier function with angular part of cubic-harmonics $Y_{lm}$.

For the system with strong spin-orbit coupling,  we use:

\begin{equation}
  \begin{split}
    F^k(i\nu_n)=&\frac{1}{4C_{k}^l}\frac{4\pi}{2k+1}\sum_{m_1,m_2,m_3,m_4}\sum_{s,s'=\pm\frac{1}{2}}\sum_{m'_1,m'_2,m'_3,m'_4}\sum_{a_1,a_2,a_3,a_4=\pm\frac{1}{2}} \\
    & \langle Y_l^{m_1}|Y_k^{m_1-m_4}Y_l^{m_4}\rangle\langle Y_l^{m_2}Y_k^{m_2-m_3}|Y_l^{m_3}\rangle \\
    &S_{m_1s,m'_1a_1}S_{m_2 s',m'_2 a_2}S^{-1}_{m_3s',m'_3a_3}S^{-1}_{m_4s,m'_4a_4}\\
    &\int d\mathbf{r}d\mathbf{r}'W_r(\mathbf{r},\mathbf{r}', R=0,i\nu_n)\\
    &W_{R=0,m'_1,a_1}^*(\mathbf{r})W_{R=0,m'_2,a_2}^*(\mathbf{r}')W_{R=0,m'_3a_3}(\mathbf{r}')W_{R=0,m'_4,a_4}(\mathbf{r})\\    
    \label{eq:slater_so}    
  \end{split}
\end{equation}
where $S_{m_1,s,m'_1,a_1}=\langle l, m_1,\frac{1}{2}, s | l+a_1, m'_1 \rangle$ and $W_{\mathbf{R}m_1a_1}(\mathbf{r})$ is a correlated Wannier function with angular part of $\Omega_{la_1m_1}$.
Parallelization of ComCoulomb is characterized by the construction of $P_{QP}^{low}$, $P_{QP}$, $W_r$, and $F^k$. The parallelization of the part to calculate $P_{QP}^{low}$ and $P_{QP}$ follows the parallelization to calculate $P_{QP}$ in FlapwMBPT code: two-dimensioal MPI-grid over $\mathbf{k}$ and $\nu_n$. The part for $W_r$ follows the parallelization to calculate $W$ in FlapwMBPT code: two-dimensioal MPI-grid over $\mathbf{k}$ and $\nu_n$.  Details can be found in Ref. \cite{kutepov_LinearizedSelfconsistentQuasiparticle_2017}. For the computation of $F^k$, the code is MPI-parallelized over $\nu_n$ with nearly linear scaling up to $N_\nu$ processors, where $N_\nu$ is the number of the bosonic frequency points treated numerically. 

\subsection{ComDC}

ComDC calculates the electron self-energy included in both \textit{ab initio} LQSGW and DMFT: the local Hartree term and the local GW term as shown in Fig.\ref{fig_functional} (c). Electron self-energy from local Hartree and local GW diagrams is as follows.
\begin{equation}
  \widetilde{\Sigma}_{DC,i,j}(i\omega_n)=\sum_{k,l=m_l'} \widetilde{G}_{l,k}(\tau=0^-)\widetilde{\mathcal{U}}_{iklj}(i\nu=0)-\sum_{k,l}\int d\tau \widetilde{G}_{l,k}(\tau)\widetilde{W}_{ikjl}(\tau)e^{i \omega_n\tau}.\label{eq:dc}
\end{equation}
where $i$,$j$,$k$ and $l$ are composite indices for both orbital and spin (spin-angular function index) in the system without (with) spin-orbit coupling. $\widetilde{\mathcal{U}}$ is constructed by using Slater's integrals in eq. \eqref{eq:slater_woso} and eq. \eqref{eq:slater_so} as well as Wannier functions associated with correlated orbitals. 
\begin{equation}
  \begin{split}
    U_{i,j,k,l}(i\nu_n)&=\sum_{\substack{m'_1m'_2,m'_3m'_4\\s_1=\pm\frac{1}{2}s_2=\pm\frac{1}{2}}}S_{i,m_1's_1}S_{j,m_2's_2}S_{k,m_3's_2}^{-1}S_{l,m_4's_1}^{-1}\\
    &\sum_{k=0}^{2l,even}\frac{4\pi}{2k+1}\langle Y_{l}^{m_1'}|Y_{k}^{q}Y_{l}^{m_4'}\rangle\langle Y_{l}^{m_2'}Y_{k}^{q}|Y_{l}^{m_3'}\rangle F^{k}(i\nu_n).    
    \label{eq:coulomb_so}
  \end{split}
\end{equation}

Here, $S_{i,m's}$ is the matrix for transformation from the spherical harmonics / spinor basis to the cubic spherical harmonics / spinor basis (the spin angular function basis) for the system without spin-orbit coupling (for the system with strong spin-orbit coupling). Meanwhile, $\widetilde{G}$ is the local Green's function and $\widetilde{W}$ is the local screened Coulomb interaction given by
\begin{equation}
  \widetilde{W}_{ikjl}(i\nu_n){=}\widetilde{\mathcal{U}}_{ikjl}(i\nu_n)+\sum_{mnpq}\allowbreak \widetilde{\mathcal{U}}_{imnl}(i\nu_n) \allowbreak \widetilde{P}_{mpqn}(i\nu_n)\allowbreak \widetilde{W}_{pkjq}(i\nu_n),\label{eq:ww}
\end{equation}
where $\widetilde{P}$ is the local polarizability and it is calculated as
\begin{equation}
  \widetilde{P}_{mpqn}(i\nu_n)\allowbreak=\int d\tau
  \widetilde{G}_{n,p}(\tau)\widetilde{G}_{q,m}(-\tau)\allowbreak e^{i\nu_n\tau}.\label{eq:pi_wso}
\end{equation}
Here we note that GW self-energy in eq. \eqref{eq:dc} can be split into Fock and correlation parts by splitting $\widetilde{W}_{ikjl}(i\nu)=\widetilde{U}_{ikjl}(i\nu=\infty)+\left(\widetilde{W}_{ikjl}(i\nu)-\widetilde{U}_{ikjl}(i\nu=\infty)\right)$ 


Parallelization of ComDC is characterized by the construction of $\widetilde{G}(\tau)$, $\widetilde{P}(i\nu_n)$, $\widetilde{W}(i\nu_n)$, $\widetilde{W}(\tau)$, $\widetilde{\Sigma}(\tau)$, and $\widetilde{\Sigma}(i\omega_n)$. For the part to calculate $\widetilde{G}(\tau)$, $\widetilde{W}(\tau)$ and $\widetilde{\Sigma}(\tau)$, the code is MPI-parallelized over $\tau$ with nearly linear scaling up to $N_{\tau}$ processors where $N_\tau$ is the number of $\tau$ mesh. For the part to calculate $\widetilde{P}(i\nu)$ and $\widetilde{W}(i\nu)$, the code is MPI-parallelized over $\nu_n$ with nearly linear scaling up to $N_{\nu}$ processors where $N_\nu$ is the number of $\nu$ mesh. For the part to calculate $\widetilde{\Sigma}(i\omega_n)$, the code is MPI-parallelized over $\omega_n$ with nearly linear scaling up to $N_{\omega}$ processors where $N_\omega$ is the number of $\omega$ mesh. Note that $N_\nu$, $N_\omega$, $N_\tau$ are linearly increasing functions of the inverse temperature and their values at 300K is typically $\sim$1,000, $\sim$4,000, and $\sim$2,000, This code scales well up to $\sim$10,000 CPU cores.

\subsection{ComLowH}\label{sec_comlowh}

ComLowH constructs non-local LQSGW Hamiltonian($H_{QP}^{nl}$) and lattice Green's function ($G$) in a fine-$\mathbf{k}$ grid using Wannier interpolation, in addition to the fermionic Weiss field ($\widetilde{\mathcal{G}}$). After Wannier-interpolating $H_{QP}$ and $f_\mathbf{k}$, non-local LQSGW Hamiltonian \cite{tomczak_QSGWDMFTElectronic_2015} is constructed in the following way. 
\begin{equation}
  \begin{split}
    H_{QP}^{nl}(\mathbf{k})=\sqrt{Z_{DC}^{-1}(\mathbf{k})} H_{QP}\sqrt{Z_{DC}^{-1}(\mathbf{k})}-f_\mathbf{k} \widetilde{\Sigma}_{DC}(\omega=0) f_\mathbf{k}^\dagger,
    \label{eq:h_qp_nl}    
  \end{split}
\end{equation}
where $Z_{DC}^{-1}(\mathbf{k})=1-f_\mathbf{k}\left({\partial{\widetilde{\Sigma}_{DC}(\omega=0)}}/{\partial{i\omega_n}}\right)f_\mathbf{k}^\dagger$. Then the lattice Green's function ($G$) is given by
\begin{equation}
  \begin{split}
    G^{-1}(\mathbf{k},i\omega_n)=i\omega_n-H_{QP}^{nl}(\mathbf{k})-f_\mathbf{k} \widetilde{\Sigma}_{imp}(i\omega_n) f_\mathbf{k}^\dagger
    \label{eq:glat_inv}
  \end{split}
\end{equation}
and the fermionic Weiss field is 
\begin{equation}
  \begin{split}
    \widetilde{\mathcal{G}}=\left(\left(\frac{1}{N_\mathbf{k}}\sum_\mathbf{k} f_\mathbf{k}^\dagger G(\mathbf{k},i\omega_n)f_\mathbf{k}\right)^{-1}+\widetilde{\Sigma}_{imp}\right)^{-1}
    \label{eq:glat}    
  \end{split}
\end{equation}

With an input of self-energy on a real frequency axis, ComLowH calculates projected DOS ($D_{I}$) of
\begin{equation}
  \begin{split}
    D_{I}(\omega)=&-\frac{1}{\pi N_k}\sum_{\mathbf{k}}Im \langle W_{I\mathbf{k}}|G(\mathbf{k},\omega)|W_{I\mathbf{k}}\rangle.
    \label{eq:PDOS}    
  \end{split}
\end{equation}
Total DOS ($D$) is
\begin{equation}
  \begin{split}
    D(\omega)=&\sum_I D_I(\omega).
    \label{eq:TDOS}    
  \end{split}
\end{equation}
The momentum resolved spectral function ($A$) is given by 
\begin{equation}
  \begin{split}
    A(\mathbf{k}, \omega)=&-\frac{1}{\pi }\sum_{I}Im \langle W_{I\mathbf{k}}|G(\mathbf{k},\omega)|W_{I\mathbf{k}}\rangle.
    \label{eq:a_omega_k}    
  \end{split}
\end{equation}


Parallelization of ComLowH is characterized by the construction of the $G^{-1}$ in eq. \eqref{eq:glat_inv}. For the computation of $G^{-1}$, the code is MPI-parallelized over $\mathbf{k}$ with nearly linear scaling up to $N_\mathbf{k}$ processors. 
Note that $N_\mathbf{k}$ for the Wannier interpolation is typically $\sim$1,000 and this code scale well up to $\sim$1,000 CPU cores.

\subsection{ComCTQMC}

\label{sec:Intro}

ComDMFT necessitates the solution of an impurity model action. In ComDMFT, hybridization-expansion continuous time quantum Monte Carlo (CTQMC) is adopted. CTQMC is a stochastic approach to obtain numerically exact solutions of an impurity model. An impurity model consists of a small interacting system, the impurity, immersed in a bath of non-interacting electrons. The action of the impurity model relevant for GW+DMFT reads

\begin{equation}
\begin{split}
\label{equ:Action}
S = &-\iint_0^\beta \sum_{ij}  c^\dagger_i(\tau) \widetilde{\mathcal{G}}_{ij}^{-1}(\tau - \tau') c_j(\tau')   d\tau d\tau' \\
&\quad\quad+\frac{1}{2}\iint_0^\beta \sum_{ijkl}  c^\dagger_i(\tau) c^\dagger_j(\tau') \widetilde{\mathcal{U}}_{ijkl}(\tau - \tau') c_k(\tau')c_l(\tau) d\tau d\tau', 
\end{split}
\end{equation}
where $c^\dagger_i$ creates an electron in the generalized orbital $i$ (which includes both spin and orbital degrees of freedom), $\beta$ is the inverse temperature, $\widetilde{\mathcal{G}}_{ij}$ is the fermionic Weiss field provided by ComLowH (Sec. \ref{sec_comlowh}) and $\widetilde{\mathcal{U}}_{ijkl}$ is the frequency dependent interaction provided by ComCoulomb (Sec. \ref{sec_comcoulomb}). 

We assume that the frequency dependent interaction is of the form
\begin{equation}
\widetilde{\mathcal{U}}_{ijkl}(i\nu_n) = \widetilde{U}_{ijkl} + F^0(i\nu_n)\delta_{il}\delta_{jk},
\end{equation}
that is, only the dynamical screening of the Slater-Condon parameter $F^0$ is taken into account, for the simplicity in the numerical algorithm based on hybridization-expansion CTQMC. The other Slater-Condon parameters, which define $\widetilde{U}_{ijkl}$, are frequency independent and approximated by their value at $i\nu_n=0$. The Weiss field can be split as 
\begin{equation}
\label{equ:WeissField}
\widetilde{\mathcal{G}}_{ij}^{-1}(i\omega_n) = (i\omega_n + \mu)\delta_{ij} - \widetilde{t}_{ij} - \widetilde{\Delta}_{ij}(i\omega_n),
\end{equation}
where $\mu$ is the chemical potential, $\widetilde{t}_{ij}$ are impurity onsite energies and $\widetilde{\Delta}_{ij}$ is the hybridization function.

Solving the impurity model eq. \eqref{equ:Action} with CTQMC starts by expanding the partition function of the impurity model in powers of the hybridization as
\begin{equation}
\label{equ:Expansion}
\begin{split}
Z&=\int \mathcal D c^\dagger \mathcal D c  e^{-S}\\
&= \sum_{k\ge 0} \frac{1}{k!} \sum_{i_1\cdots i_k} \sum_{j_1\cdots j_k} \iint_0^\beta d\tau_1d\tau_1'\cdots\iint_0^\beta d\tau_k d\tau_k'\\
&\times \text{Tr}[e^{-\beta \hat{H}_\text{loc}} \text{T}_\tau \hat{c}_{j_1}(\tau_1')\hat{c}_{i_1}^\dagger (\tau_1)\cdots \hat{c}_{j_k}(\tau_k')\hat{c}_{i_k}^\dagger (\tau_k)]\\
&\times e^{\frac{1}{2}\textstyle{\sum}_{1\leq n,m\leq k}K(\tau_n-\tau_m)-K(\tau_n'-\tau_m)-K(\tau_n-\tau_m')+K(\tau_n'-\tau_m')}\\
&\times \underset{1\leq n,m\leq k}{\mathrm{Det}} \widetilde{\Delta}_{i_nj_m}(\tau_n-\tau'_m).
\end{split}
\end{equation}
Here we defined 
\begin{equation}
K(\tau) = \frac{F^{0,\text{ret}}(i\nu_n = 0)}{2\beta} |\tau|(|\tau| - \beta)+ \sum_{n \ne 0} \frac{F^{0,\text{ret}}(i\nu_n)}{\beta(i\nu_n)^2}e^{-i\nu_n\tau} 
\end{equation}
and 
\begin{equation}
\hat{H}_\text{loc}=\sum_{ij} \hat{c}^\dagger_i( \widetilde{t}_{ij} - \mu^\text{scr}\delta_{ij}) \hat{c}_j + \frac{1}{2}\sum_{ijkl}\hat{c}^\dagger_i \hat{c}^\dagger_j\widetilde{U}_{ijkl}^{\text{scr}}\hat{c}_k \hat{c}_l, 
\end{equation}
where $F^{0,\text{ret}}(i\nu_n) = F^0(i\nu_n) - F^0(i\nu_n = \infty)$, $\mu^\text{scr} = \mu - \frac{1}{2}F^{0,\text{ret}}(i\nu_n = 0)$ and $\widetilde{U}_{ijkl}^{\text{scr}} = \widetilde{U}_{ijkl} + F^0(i\nu_n = 0)\delta_{il}\delta_{jk}$.
Sampling the expansion eq.\eqref{equ:Expansion} with a Metropolis-Hasting Markov-Chain algorithm then yields estimates 
of observables, see refs.\cite{haule_QuantumMonteCarlo_2007,gull_ContinuoustimeMonteCarlo_2011,werner_DynamicalScreeningCorrelated_2010} for a detailed description. 

The most important observable within DMFT is the Green function on the impurity
\begin{equation}
\widetilde{G}_{ij}(\tau) = -Z^{-1}\int \mathcal D c^\dagger \mathcal D c  e^{-S} c_{i}(\tau)c^\dagger_{j},
\end{equation}
from which the self-energy 
\begin{equation}
\widetilde{\Sigma}_{ij}(i\omega_n) = \widetilde{\mathcal{G}}_{ij}^{-1}(i\omega_n) - \widetilde{G}_{ij}^{-1}(i\omega_n) 
\end{equation}
can be extracted.

ComCTQMC is parallelized by assigning one Markov chain to every MPI process, and one MPI process to every CPU core of a parallel machine. Since the Markov chains are independent, there is essentially no communication between the MPI processes, an parallel performance is ideal, as long as the thermalization time of the Markov chains is negligible to the sampling time. To ensure this, we take advantage of the iterative method to solve the self-consistent DMFT equation and use the state of the Markov chains from previous DMFT iterations as starting point. This makes thermalization time negligible close to self-consistency, and the thermalization time needed in the first few iterations is small when compared to the time needed for converging the self-consistent DMFT equation

\section{Computational Layout for charge self-consistent LDA+DMFT}\label{sec_lda_dmft}
\begin{figure}
  \centering
  \includegraphics[width=1.0\columnwidth]{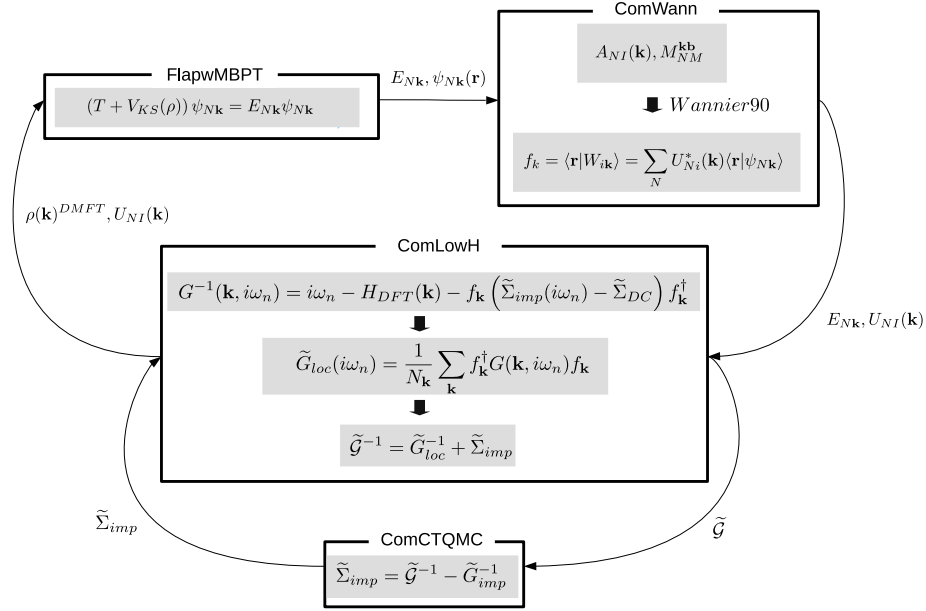}
  \caption{charge self-consistent LDA+DMFT flowchart and data exchange between programs (FlapwMBPT, ComWann, ComLowH, and ComCTQMC) in ComDMFT. For the meaning of each symbol, please see table \ref{tab:symbol}. First, a LDA Hamiltonian is constructed by using FlapwMBPT \cite{kutepov_LinearizedSelfconsistentQuasiparticle_2017}. In ComWann, with LDA energies and wavefunctions, initial trial rotation matrix ($A_{NI}(\mathbf{k})$) and overlap matrix ($M_{NM}^{\mathbf{k}\mathbf{b}}$) are calculated to construct Wannier functions and their rotation matrices ($U_{NI}(\mathbf{k})$) by using Wannier90 package \cite{mostofi_Wannier90ToolObtaining_2008}. We solve DMFT self-consistent loop by using ComLowH and ComCTQMC. ComLowH Wannier-interpolates LDA bandstructure and construct Lattice Green's function ($G$) and fermionic Weiss field ($\widetilde{\mathcal{G}}$) by compensating double-counting self-energy and embedding impurity self-energy ($\widetilde{\Sigma}_{imp}$). ComCTQMC calculates impurity self-energy. Finally, a density matrix ($\rho^{DMFT}(\mathbf{k})$) in the low-energy Hilbert space are embedded into the full density matrix to update electron density. Slater's integrals ($F^k$) and electron occupation associated with the correlated orbitals (for the nominal double counting energy scheme \cite{haule_DynamicalMeanfieldTheory_2010,haule_CovalencyTransitionmetalOxides_2014}) are supposed to be provided by a user}  
  \label{fig_flow_lda_dmft}
\end{figure}
In ComDMFT, charge self-consistent LDA+DMFT methodology is implemented, enabling multiple methods for the electronic structure of CES in one platform. In addition to the crystal structure and the choice of correlated orbitals, the LDA+DMFT method requires the interaction strength $\widetilde{\mathcal{U}}$, which can parameterized by Slater's integrals, and a double counting correction. 
The double counting correction, is expressed in terms of the electron occupation associated with the correlated orbitals. As a default, we take the the nominal double counting 
scheme \cite{haule_DynamicalMeanfieldTheory_2010,haule_CovalencyTransitionmetalOxides_2014}). The nomainal occupation along with Slater's integrals are to be provided by a user for the charge self-consistent LDA+DMFT calculations. Charge self-consistent LDA+DMFT calculations can be broken down into four steps as shown in Fig. \ref{fig_flow_lda_dmft}: i) construction of a LDA Hamiltonian by FlapwMBPT \cite{kutepov_LinearizedSelfconsistentQuasiparticle_2017}, ii) construction of the atom-centered local basis set spanning the low-energy Hilbert space by ComWann utilizing Wannier90 package \cite{mostofi_Wannier90ToolObtaining_2008}, iii) Wannier interpolation of the LDA Hamiltonian and solving the DMFT self-consistent equation by ComLowH and ComCTQMC, and iv) updating electron density. 

The electron density is updated in the following way. The local density is defined as

\begin{equation}
  \begin{split}
    \rho(\mathbf{r})=\frac{1}{N_\mathbf{k}}\sum_{\mathbf{k},N,M}\langle \mathbf{r}|\psi_{N\mathbf{k}}\rangle \rho_{NM}(\mathbf{k}) \langle \psi_{M\mathbf{k}}|\mathbf{r}\rangle
  \end{split}
  \label{eq:density}  
\end{equation}
where $\rho_{NM}(\mathbf{k})$ is a density matrix spanned by band eigenstates. To update charge density, low-energy density matrix from DMFT self-consistent equation solution is embedded to the density matrix in eq. \eqref{eq:density} \begin{equation}
  \begin{split}
    \rho_{NM}(\mathbf{k})&=\rho_{N}^{LDA}\delta_{NM}
    +\sum_{IJ}\langle \psi_{N\mathbf{k}}|W_{I\mathbf{k}}\rangle \\
    &\left(\rho_{IJ}^{DMFT}(\mathbf{k}) -\sum_{N'}\langle W_{I\mathbf{k}}|\psi_{N'\mathbf{k}}\rangle\rho_{N'}^{LDA}\langle \psi_{N'\mathbf{k}}|W_{J\mathbf{k}}\rangle\right) \langle W_{J\mathbf{k}}|\psi_{M\mathbf{k}}\rangle
  \end{split}
  \label{eq:density_matrix}  
\end{equation}
where $\rho_N^{LDA}(\mathbf{k})$ is the density matrix within density functional theory and $\rho_{IJ}^{DMFT}(\mathbf{k})$ is the density matrix from DMFT self-consistent equation solution in the low-energy Hilbert space. Ef.\eqref{eq:density} and Ef. \eqref{eq:density_matrix} are implemented in FlapwMBPT.

\section{Results}\label{sec_results}

To benchmark the performance of ComDMFT, we calculate the electronic structure of MnO, NiO, and FeSe at T=300K within \textit{ab initio} LQSGW+DMFT. These are protopypical CES and have been studied intensively within DFT+DMFT\cite{kunes_CollapseMagneticMoment_2008,kunes_LocalCorrelationsHole_2007a,kunes_NiOCorrelatedBand_2007,ren_MathrmLDAMathrmDMFTComputation_2006,karolak_DoubleCountingLDA_2010a,mandal_StrongPressuredependentElectronphonon_2014,yin_SpinDynamicsOrbitalantiphase_2014,aichhorn_TheoreticalEvidenceStrong_2010,watson_FormationHubbardlikeBands_2017,leonov_CorrelationDrivenTopologicalFermi_2015,haule_ForcesStructuralOptimizations_2016,skornyakov_CorrelationStrengthLifshitz_2018,yin_KineticFrustrationNature_2011,zhang_DFTDMFTCalculations_2017} and \textit{ab initio} GW\cite{faleev_AllElectronSelfConsistentGW_2004,kobayashi_GWApproximationLSDA_2008,patterson_ComparisonHybridDensity_2006,li_QuasiparticleEnergyBands_2005, aryasetiawan_ElectronicStructureNiO_1995,tomczak_ManyBodyEffectsIron_2012}. MnO and NiO are paramagnetic Mott insulators with rocksalt crystal structure. FeSe is a paramagnetic Hund's metal which shows nematic order and unconventional superconductivity at low temperature. We compare \textit{ab initio} LQSGW+DMFT results with experiments. Charge self-consistent LDA+DMFT results obtained by using ComDMFT are also shown for comparison. In addition to the density of states and crystal momentum resolved spectral function of three CES, some of the important physical quantities are shown for the demonstration of ComDMFT. For NiO and MnO in their rocksalt crystal structure, lattice constants of  4.194 $\AA$ \cite{bartel_ExchangeStrictionNiO_1971} and 4.445 $\AA$ \cite{johnston_StudyLixMn1x_1956} are used, respectively. For FeSe, the experimentally-determined crystal structure in P4/nmm space group \cite{phelan_NeutronScatteringMeasurements_2009} is used. 

\subsection{Wannier functions and interpolated Wannier bandstructure of MnO}
\begin{figure}
  \centering
  \includegraphics[width=1.0\columnwidth]{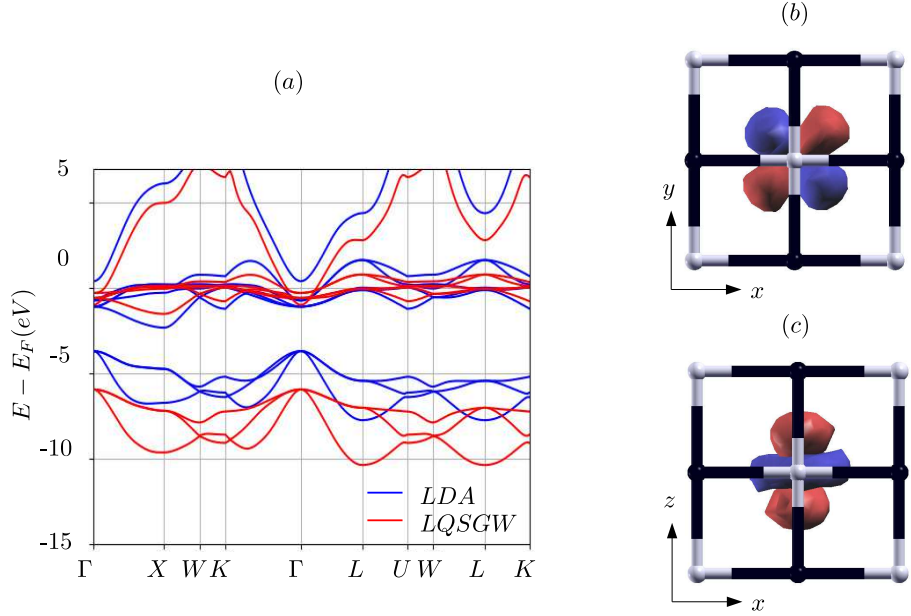}
  \caption{(color online) (a) Wannier interpolated MnO bandstructures within non spin-polarized LDA and LQSGW. (b) Mn-$d_{xy}$ and (c) Mn-$d_{z^2}$ Wannier functions in a FCC conventional unitcell. White and black spheres represent O and Mn atoms, respectively}
  \label{fig_band_wan}
\end{figure}

Figure \ref{fig_band_wan} (a) shows the Wannier-interpolated electronic bandstructure of MnO within non spin-polarized LDA and LQSGW. As demonstrated in this figure, conventional electronic structure methods based on one-particle picture, in their non spin-polarized versions, fail to reproduce an insulating gap in MnO. Although non spin-polarized LQSGW shows bandwidth narrowing for the five bands near the Fermi level, it still predicts that MnO is a metal. To define five Mn-$d$ orbitals as correlated orbitals, Wannier functions for Mn-$s$, Mn-$p$ Mn-$d$, and O-$p$ orbitals are constructed in a frozen energy window of $\text{-}15 eV <E-E_f < 7eV$. Figure \ref{fig_band_wan} (b) and (c) show Mn-$d_{xy}$ and Mn-$d_{z^2}$ Wannier functions in a conventional FCC unit cell, respectively. They are centered exactly at a Mn atom and show expected angular distributions of the wavefunctions. The square root of the spread of the Mn-$t_{2g}$ and Mn-$e_{g}$ orbitals are ,$\sqrt{\langle\mathbf{r}^2-\langle\mathbf{r}\rangle_{\mathbf{R}=0,Mn\text{-}t_{2g}}^2\rangle_{\mathbf{R}=0,Mn\text{-}t_{2g}}}=0.69\AA$ and $\sqrt{\langle\mathbf{r}^2-\langle\mathbf{r}\rangle_{\mathbf{R}=0,Mn\text{-}e_{g}}^2\rangle_{\mathbf{R}=0,Mn\text{-}e_{g}}}=0.62\AA$, respectively. These spreadings are much smaller than the interatomic distance between Mn and O of 2.22 $\AA$. Spin-integrated electron occupations in Mn-$d$ orbitals within LQSGW are $\langle n_{Mn\text{-}{d_{xy}}}\rangle =\langle n_{Mn\text{-}{d_{yz}}}\rangle=\langle n_{Mn\text{-}{d_{zx}}}\rangle=1.4$ and $\langle n_{Mn\text{-}{d_{z^2}}}\rangle =\langle n_{Mn\text{-}{d_{x^2-y^2}}}\rangle=\langle n_{Mn-{zx}}\rangle=0.4$, where $n_I$ is the density operator in an orbital $I$. These values are far from the value at half-filling and hinder a paramagnetic Mott gap opening. 

\subsection{Coulomb interaction tensor and Slater's integrals associated with Mn-$d$ orbitals in MnO}
\begin{figure}
  \centering
  \includegraphics[width=1.0\columnwidth]{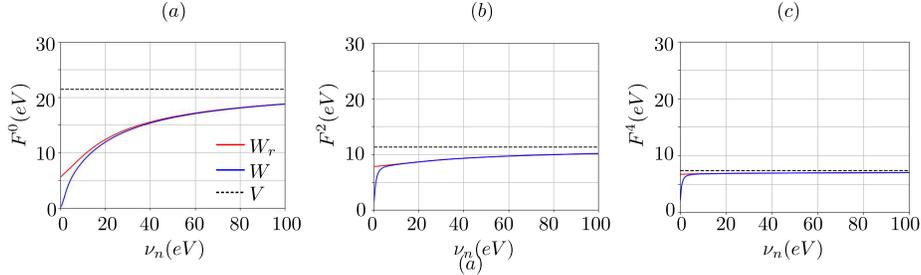}
  \caption{(color online) Slater's integrals associated with Mn-$d$ orbitals: (a) $F^0$, (b) $F^2$, and (c) $F^4$. Red and blue colored lines correspond to the Slater's integrals of partially screened Coulomb interaction $W_r$ and fully screened Coulomb interaction $W$, respectively. Black dashed lines are for bare Coulomb interaction. For DMFT calculation, only dynamical screening associated $F^0$ has been considered. }
  \label{fig_slaters}
\end{figure}

Figure \ref{fig_slaters} shows  Slater's integrals of partially-screened Coulomb interactions in eq. \eqref{eq:slater_woso}. Slater's integrals of bare Coulomb interaction and fully-screened Coulomb interaction are shown for comparison. These quantities are obtained by replacing $W_r$ with $V$ and $W$ in in eq. \eqref{eq:slater_woso}, respectively. By excluding polarizability between the five correlated bands, dielectric screening are suppressed and Slater's integrals of partially screened Coulomb interaction are larger than those of fully-screened Coulomb interactions at low-energy. In addition, monopole integral ($F^0$) shows much stronger frequency dependence than $F^2$ and $F^4$. To illustrate, $F^0$ of partially-screened Coulomb interaction increases from 5.6eV to 21.5eV but $F^4$ from 6.9eV to 7.3eV.
  
\subsection{Impurity self-energies associated with Mn-$d$ orbitals}
\begin{figure}
  \centering
  \includegraphics[width=1.0\columnwidth]{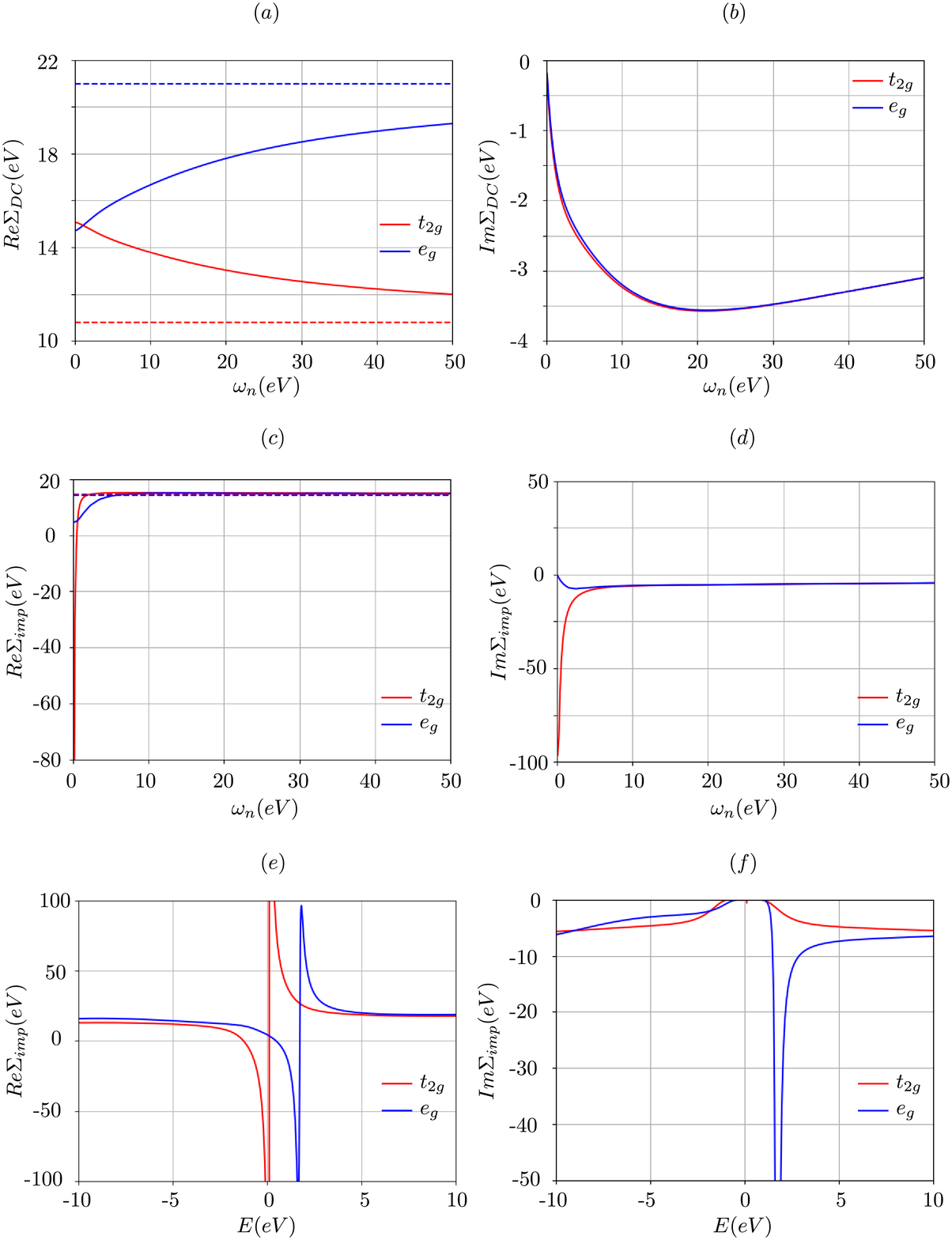}
  \caption{(color online) (a) real and (b) imaginary parts of spin and orbital resolved local-GW self-energies associated with a Mn-$t_{2g}$ (red) and a Mn-$e_{g}$ (blue) orbital on the imaginary frequency axis. Dotted lines show their high-frequency limit. (c) real and (d) imaginary part of spin and orbital resolved DMFT self-energies associated with a Mn-$t_{2g}$ (red) and a Mn-$e_{g}$ (blue) orbital on the imaginary frequency axis. (e) real and (f) imaginary parts of spin and orbital resolved impurity self-energies associated with a Mn-$t_{2g}$ (red) and a Mn-$e_{g}$ (blue) orbital on the real frequency axis.}
  \label{fig_sig}
\end{figure}

Figure \ref{fig_sig} (a) and (b) show double-counted electronic self-energies within the local-GW approximation associated with five Mn-$d$ orbitals on the imaginary frequency axis. Red and blue lines are for Mn-$t_{2g}$ and Mn-$e_{g}$ orbitals, respectively. Both real and imaginary parts of self-energies do not show divergent behaviors and their imaginary parts are even linear in imaginary frequency near the Fermi level. However, if all Feynman diagrams associated with five Mn-$d$ orbitals are summed, self-energies show qualitatively different behaviors. Figure \ref{fig_sig} (c) and (d) show impurity self-energies from ComCTQMC on the imaginary frequency axis. In contrast to electronic self-energies within local GW approximation, both real and imaginary parts of DMFT self-energy on the imaginary frequency axis show divergent behaviors near the Fermi level. In addition, spin-integrated electron occupations in Mn-$d$ orbitals within DMFT are $\langle n_{Mn\text{-}{d_{xy}}}\rangle =\langle n_{Mn\text{-}{d_{yz}}}\rangle=\langle n_{Mn\text{-}{d_{zx}}}\rangle=1.0$ and $\langle n_{Mn\text{-}{d_{z^2}}}\rangle =\langle n_{Mn\text{-}{d_{x^2-y^2}}}\rangle=\langle n_{Mn-{zx}}\rangle=1.04$, which are much closer to the value at half-filling than those within LQSGW approximation. Analytical continuation by using maximum entropy method \cite{jarrell_BayesianInferenceAnalytic_1996a} implemented in EDMFTF package \cite{_EmbeddedDMFTFunctional_} results in divergent electronic self-energies more clearly as shown in Fig. \ref{fig_sig} (e) and (f). Electronic self-energy for both Mn-$t_{2g}$ and Mn-$e_{g}$ has a pole near the Fermi level, inducing a paramagnetic Mott gap in MnO. Here we note that the analytical continuation code is not in the ComDMFT package.

\subsection{Hybridization functions associated with Mn-$d$ orbitals}

\begin{figure}
  \centering
  \includegraphics[width=1.0\columnwidth]{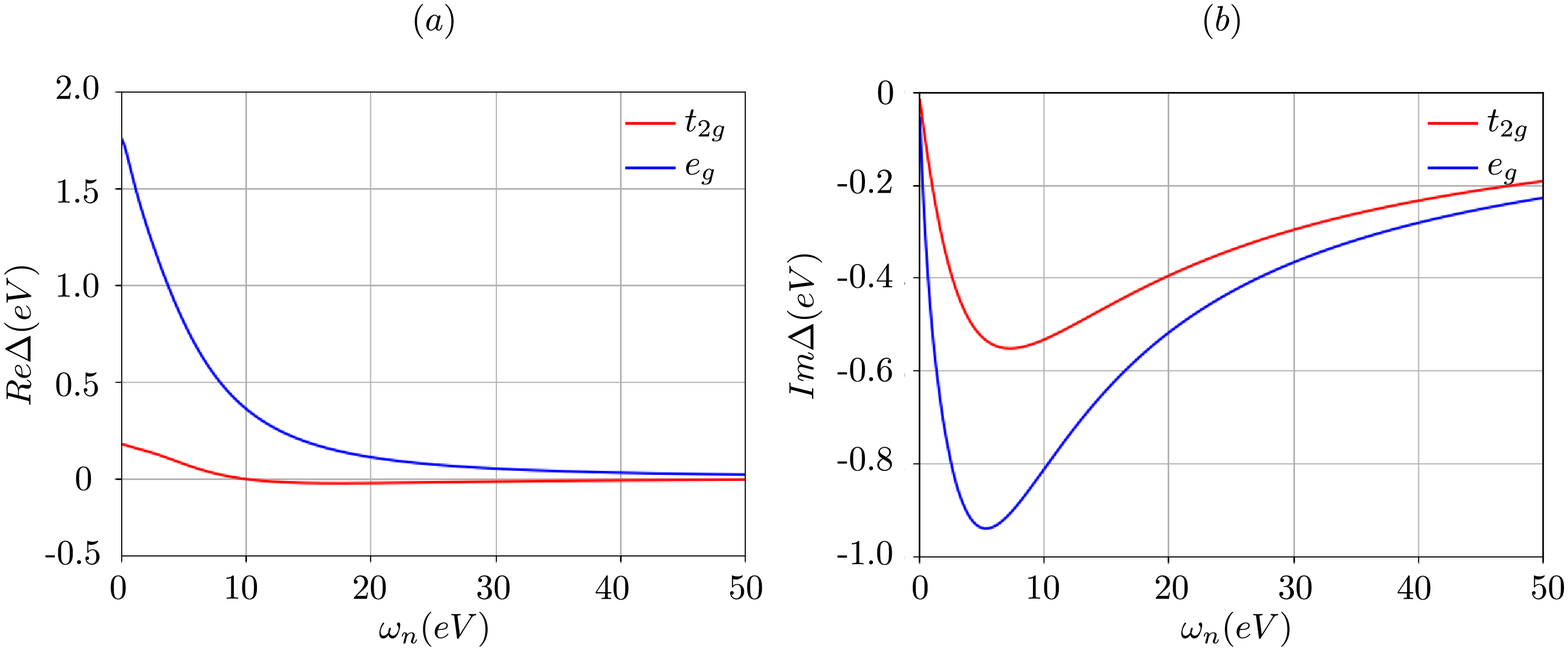}
  \caption{(color online) (a) real and (b) imaginary parts of spin and orbital resolved hybridization functions associated with a Mn-$t_{2g}$ (red) orbital and a Mn-$e_g$ (blue) orbital on the imaginary frequency axis.}
  \label{fig_delta}
\end{figure}

Figure \ref{fig_delta} shows hybridization functions associated with five Mn-$d$ orbitals on the imaginary frequency axis. Red and blue lines are for Mn-$t_{2g}$ and Mn-$e_{g}$ orbitals, respectively. A Mn-$t_{2g}$ orbital experiences less hybridization with the rest of the electrons than a Mn-$e_{g}$ orbital. Imaginary part of the hybridization function approaches to zero at the Fermi level, implying the opening of an energy gap in MnO. 

\subsection{The density of states and spectral functions of MnO}
\begin{figure}
  \centering
  \includegraphics[width=1.0\columnwidth]{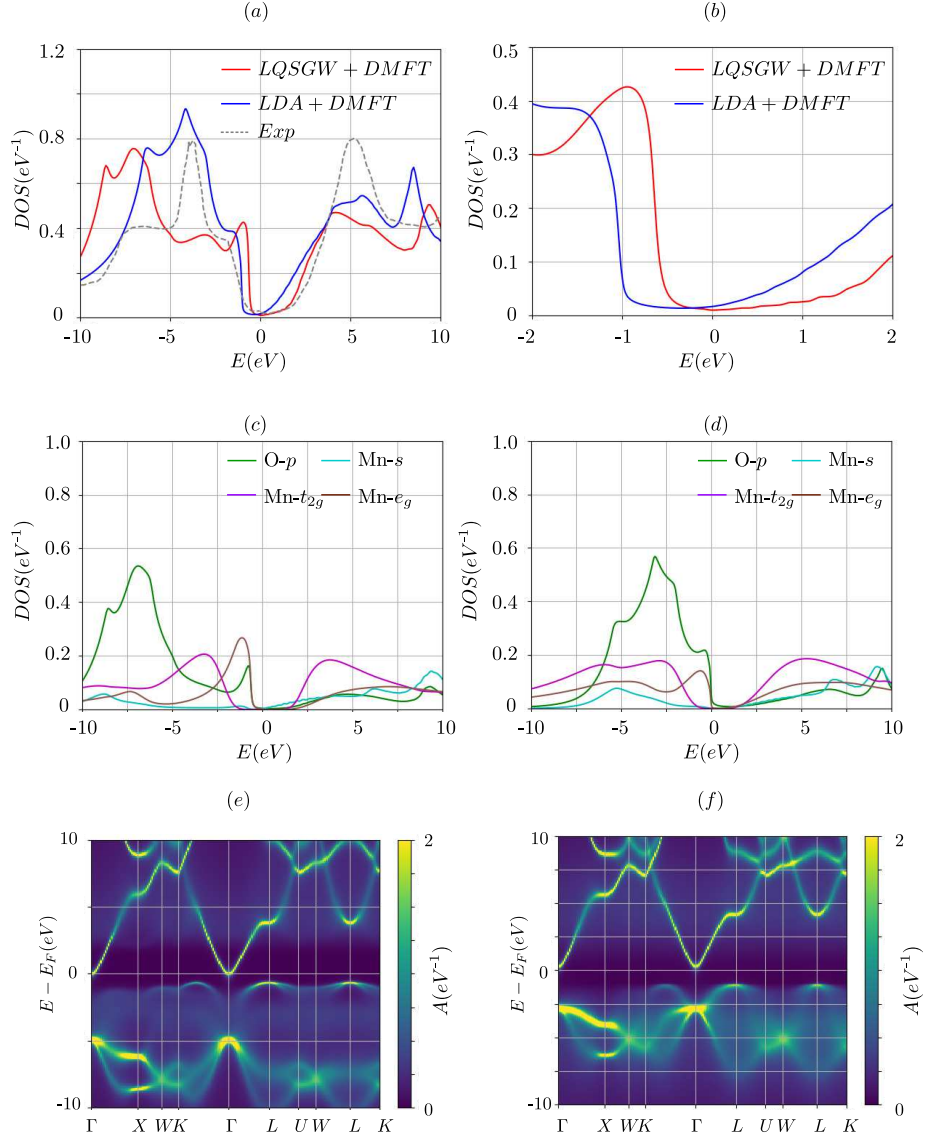}
  \caption{(color online) (a) MnO density of states and (b) its zoom-in view. Red and blue lines show the total density of states within \textit{ab initio} LQSGW+DMFT and charge self-consistent LDA+DMFT. Gray dashed lines are from photoemission spectroscopy and Bremsstrahlung Isochromat Spectroscopy \cite{kurmaev_OxygenXrayEmission_2008,vanelp_ElectronicStructureMnO_1991}. (c) the projected density of states to O-$p$, Mn-$s$, Mn-$t_{2g}$, and Mn-$e_{g}$ orbitals within \textit{ab initio} LQSGW+DMFT, marked by green, cyan, purple, and brown colors, respectively. (d) the projected density of states to O-$p$, Ni-$s$, Ni-$t_{2g}$, and Ni-$e_{g}$ orbitals within charge self-consistent LDA+DMFT, marked by green, cyan, purple, and brown colors, respectively. (e) spectral function along a high symmetry line in the first Brillouin zone within \textit{ab initio} LQSGW+DMFT. (f) spectral function along a high symmetry line in the first Brillouin zone within charge self-consistent LDA+DMFT}
  \label{fig_spectra_mno}
\end{figure}
In this subsection, total DOS, projected DOS and crystal momentum resolved spectral function within \textit{ab initio} LQSGW+DMFT are presented. For comparison, charge self-consistent LDA+DMFT calculation is performed by using ComDMFT. For the LDA+DMFT, five Mn-$d$ orbitals are considered as correlated orbitals. $F^0=9.0eV$, $F^2=9.8eV$ and $F^4=6.1eV$ are chosen to construct Coulomb interaction tensor associated Mn-$d$ orbitals to define a impurity problem. These values are chosen to match experiments and consistent with the values in the EDMFTF LDA+DMFT database\cite{_Kristjandatabase_}. These correspond to $U=9.0eV$ and $J=1.14eV$. To define five Mn-$d$ orbitals, Wannier functions for Mn-$s$, Mn-$p$ Mn-$d$, and O-$p$ orbitals are constructed in a frozen energy window of $\text{-}10 eV <E-E_f < 10eV$ at every charge self-consistency loop. For the double-counting energy, the nominal double-counting scheme \cite{haule_DynamicalMeanfieldTheory_2010,haule_CovalencyTransitionmetalOxides_2014} is used with the $d$ orbital occupancy of 5.0. LDA+DMFT results from ComDMFT are confirmed to reproduce the results from EDMFTF\cite{haule_DynamicalMeanfieldTheory_2010,_EmbeddedDMFTFunctional_}. 

Figure \ref{fig_spectra_mno} (a) show MnO density of states within charge self-consistent LDA+DMFT and \textit{ab initio} LQSGW+DMFT. For comparison, photoemission spectroscopy and bremsstrahlung isochromat spectroscopy results \cite{kurmaev_OxygenXrayEmission_2008,vanelp_ElectronicStructureMnO_1991} are reproduced. Both charge self-consistent LDA+DMFT and \textit{ab initio} LQSGW+DMFT open an insulating gap, which can be clearly seen in a zoom-in view in Figure \ref{fig_spectra_mno} (b). They reproduce experimentally observed four peak structure at 5eV, -2eV, -4eV and -7eV from the Fermi level reasonably well. The projected density of state calculation within LQSGW+DMFT shown in Fig. \ref{fig_spectra_mno} (c) attributes each peak to Mn-$t_{2g}$, Mn-$e_{g}$, Mn-$t_{2g}$, and O-$p$ orbitals, respectively. In contrast, all peaks below the Fermi level are dominated by O-$p$ orbitals within charge self-consistent LDA+DMFT, as shown in Fig. \ref{fig_spectra_mno} (d). Substantial contribution of O-$p$ orbitals to the top of the valence band shows strong-hybridization between O-$p$ and Mn-$e_g$, consistent with Zhang-Rice multiplet picture. This strong hybridization gives rise to sharp spectral features at the top of the valence bands in the crystal Momentum resolved spectral function. Fig. \ref{fig_spectra_mno} (e) and (f) show momentum resolved spectral functions within \textit{ab initio} LQSGW+DMFT and charge self-consistent LDA+DMFT. Above the Fermi level, band structure and the position of the Hubbard bands are quite similar. However, below the Fermi level, they show differences. Although Zhang-Rice multiplet peaks at the top of the valence bands are sharp and become weaker near the $\Gamma$ point from both methods, O-$p$ bands (bands at $E_F\text{-}5eV$ at $\Gamma$ point) are at a lower energy within LQSGW+DMFT than charge self-consistent LDA+DMFT.

\begin{figure}
  \centering
  \includegraphics[width=1.0\columnwidth]{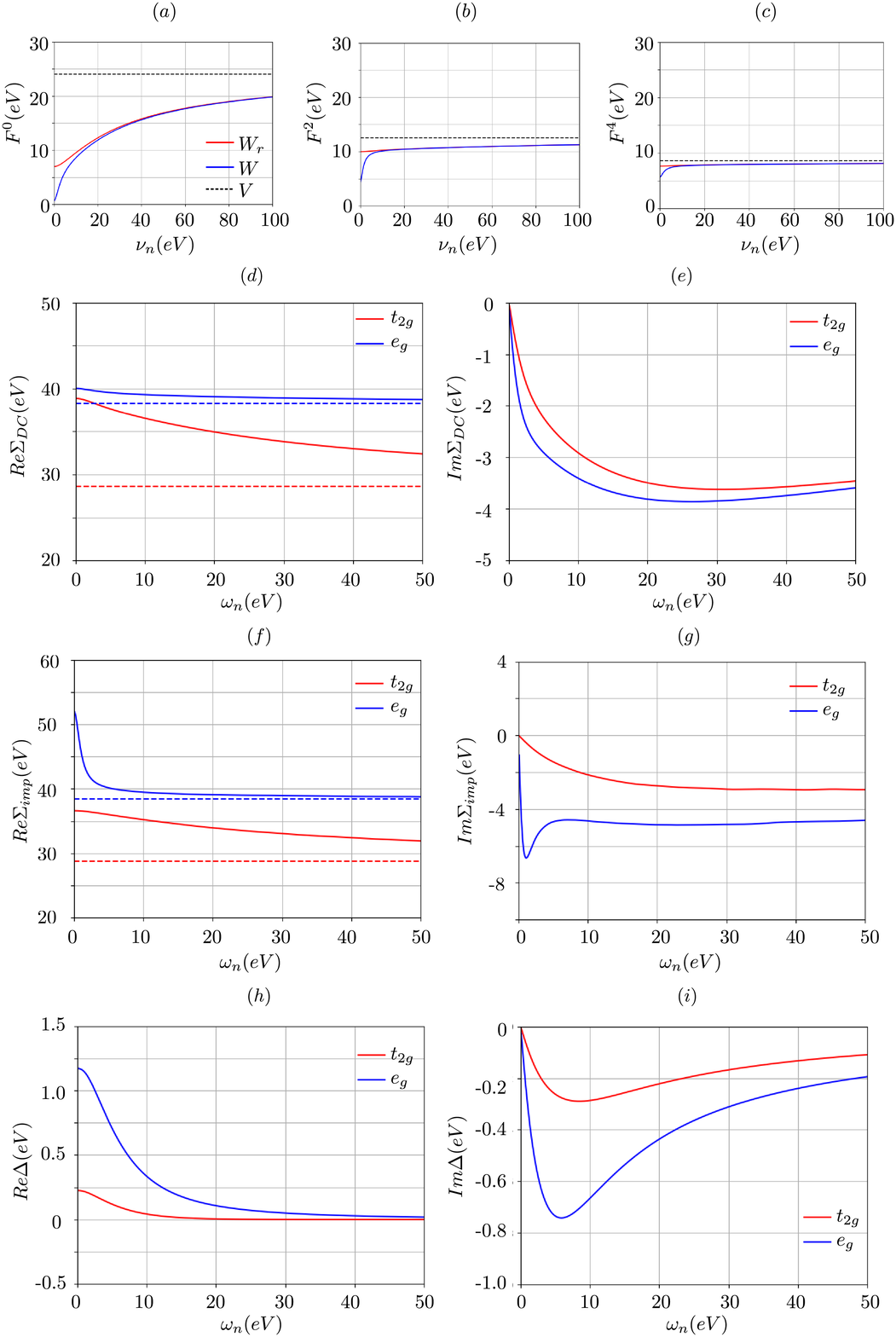}
  \caption{(color online) Slater's integrals associated with Ni-$d$ orbitals: (a) $F^0$, (b) $F^2$, and (c) $F^4$. Red and blue colored lines correspond to the Slater's integrals of partially screened Coulomb interaction $W_r$ and fully screened Coulomb interaction $W$, respectively. Black dashed lines are for bare Coulomb interaction. For DMFT calculation, only dynamical screening associated $F^0$ has been considered. (d) real and (e) imaginary parts of spin and orbital resolved local-GW self-energies associated with a Ni-$t_{2g}$ (red) and a Ni-$e_{g}$ (blue) orbital. Dotted lines show their high-frequency limit. (e) real and (f) imaginary part of spin and orbital resolved DMFT self-energies associated with a Ni-$t_{2g}$ (red) and a Ni-$e_{g}$ (blue) orbital. (g) real and (h) imaginary parts of spin and orbital resolved hybridization function associated with a Ni-$t_{2g}$ (red) and a Ni-$e_{g}$ (blue) orbitals.}
  \label{fig_nio_imp}
\end{figure}

\subsection{The density of states and spectral functions of NiO}
In this subsection, total DOS, projected DOS and crystal momentum resolved spectral function of NiO are presented. For the choice of correlated orbitals, five Ni-$d$ orbitals are chosen and Wannier functions for Ni-$s$, Ni-$p$ Ni-$d$, and O-$p$ orbitals are constructed in a frozen energy window of $\text{-}10 eV <E\text{-}E_f < 10eV$. Physical quantities associated with Ni-$3d$ orbitals for the impurity problem are shown in fig. \ref{fig_nio_imp}. For comparison, charge self-consistent LDA+DMFT calculation is performed by using ComDMFT. For the LDA+DMFT, five Ni-$d$ orbitals are considered as correlated orbitals. To define five Ni-$d$ orbitals, Wannier functions for Ni-$s$, Ni-$p$ Ni-$d$, and O-$p$ orbitals are constructed in a frozen energy window of $\text{-}10 eV <E\text{-}E_f< 10eV$ at every charge self-consistency loop. $F^0=10.0eV$, $F^2=7.8eV$ and $F^4=4.8eV$ are chosen to construct Coulomb interaction tensor associated Ni-$d$ orbitals, corresponding to $U=10.0eV$ and $J=0.9eV$. These values are chosen to match experiments and consistent with the values in the EDMFTF LDA+DMFT database\cite{_Kristjandatabase_}. For the double-counting energy, nominal double-counting scheme \cite{haule_DynamicalMeanfieldTheory_2010,haule_CovalencyTransitionmetalOxides_2014} is used with $d$ orbital occupancy of 8.0. LDA+DMFT results from ComDMFT are confirmed to reproduce the results from EDMFTF\cite{haule_DynamicalMeanfieldTheory_2010,_EmbeddedDMFTFunctional_}.

\begin{figure}
  \centering
  \includegraphics[width=1.0\columnwidth]{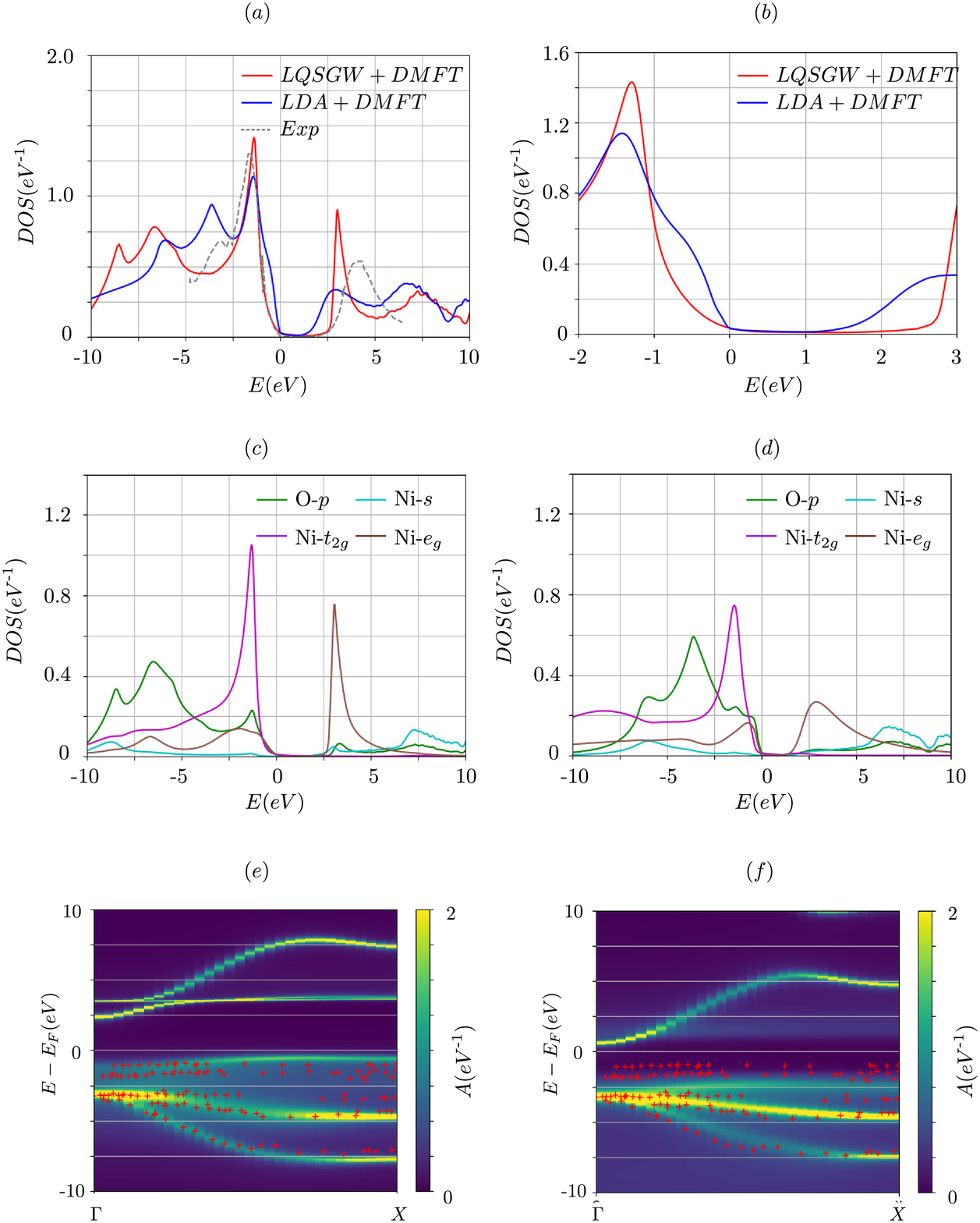}

  \caption{(color online) (a) NiO density of states and (b) its zoom-in view. Red and blue lines show the total density of states within \textit{ab initio} LQSGW+DMFT and charge self-consistent LDA+DMFT. Gray dashed lines are from photoemission spectroscopy and Bremsstrahlung Isochromat Spectroscopy \cite{sawatzky_MagnitudeOriginBand_1984}. (c) the projected density of states to O-$p$, Ni-$s$, Ni-$t_{2g}$, and Ni-$e_{g}$ orbitals within \textit{ab initio} LQSGW+DMFT, marked by green, cyan, purple, and brown colors, respectively. (d) the projected density of states to O-$p$, Ni-$s$, Ni-$t_{2g}$, and Ni-$e_{g}$ orbitals within charge self-consistent LDA+DMFT, marked by green, cyan, purple, and brown colors, respectively. (e) spectral function along $\Gamma$-$X$ in the first Brillouin zone within \textit{ab initio} LQSGW+DMFT. (f) spectral function along $\Gamma$-$X$ in the first Brillouin zone within charge self-consistent LDA+DMFT. Red plus symbols in (e) and (f) are from angle-resolved photoemission spectroscopy data\cite{shen_ElectronicStructureNiO_1991}.}
  \label{fig_spectra_nio}
\end{figure}

Figure \ref{fig_spectra_nio} (a) shows NiO total DOS within \textit{ab initio} LQSGW+DMFT. For comparison, photoemission spectroscopy and bremsstrahlung isochromat spectroscopy results \cite{sawatzky_MagnitudeOriginBand_1984} are reproduced and marked by gray dashed lines. Total DOS within charge self-consistent LDA+DMFT are marked by blue lines. Both charge self-consistent LDA+DMFT and \textit{ab initio} LQSGW+DMFT open an insulating gap, which can be clearly seen in a zoom-in view in Figure \ref{fig_spectra_nio} (b). They reproduce experimentally observed two peak structure at -2eV, and 4eV from the Fermi level reasonably well. LQSGW+DMFT projected density of state shown in Fig. \ref{fig_spectra_nio} (c) attributes each peak to Ni-$t_{2g}$ and Ni-$e_{g}$ orbitals, respectively in agreement with charge self-consistent LDA+DMFT shown in Fig. \ref{fig_spectra_nio} (d). Although the main peak below the Fermi level is dominated by Ni-$t_{2g}$, the very top of the valence bands is dominated by Ni-$e_{g}$. The subpeak at $E_F\text{-}4eV$ is hardly reproduced within LQSGW+DMFT, although LDA+DMFT give rise to a strong O-$p$ peak at slightly lower energy. There is small enhancement of the Ni-$e_{g}$ weight around $E_F\text{-}3eV$ within LQSGW+DMFT but it is too small to give rise to a subpeak. There are other LDA+DMFT results that peaks at $E_F\text{-}2eV$ and $E_F\text{-}4eV$ are originated from Ni-$d$ orbitals \cite{ren_MathrmLDAMathrmDMFTComputation_2006,kunes_NiOCorrelatedBand_2007,kunes_LocalCorrelationsHole_2007a}, but the understanding the nature of the subpeak at $E_F\text{-}4eV$ remains a puzzle to be solved \cite{kuo_ChallengesExperimentElectronic_2017}.

Fig. \ref{fig_spectra_nio} (e) and (f) show momentum resolved spectral functions within LQSGW+DMFT as well as charge self-consistent LDA+DMFT. For comparison, data from angle-resolved photoemission spectroscopy is reproduced and marked by red plus symbols\cite{shen_ElectronicStructureNiO_1991}. Above the Fermi level, Ni-$s$ bands from two different methods are qualitatively similar, but upper Hubbard bands are much sharper within LQSGW+DMFT than LDA+DMFT. Below the Fermi level, there are noticeable differences. Although the top of the valence bands is well reproduced within LQSGW+DMFT, this band is too low in energy along $\Gamma$-$X$ line within charge self-consistent LDA+DMFT. Here we note that there is also a report claiming a good agreement between LDA+DMFT and spectral function along $\Gamma$-$X$ line \cite{kunes_NiOCorrelatedBand_2007}. This discrepancy between different LDA+DMFT calculations might be due to a different choice of double counting scheme, the level of charge self-consistency and no fine adjustment on $U$ and $J$ values in our case. The influence of all these factors can be explored building on this platform.

\begin{figure}
  \centering
  \includegraphics[width=1.0\columnwidth]{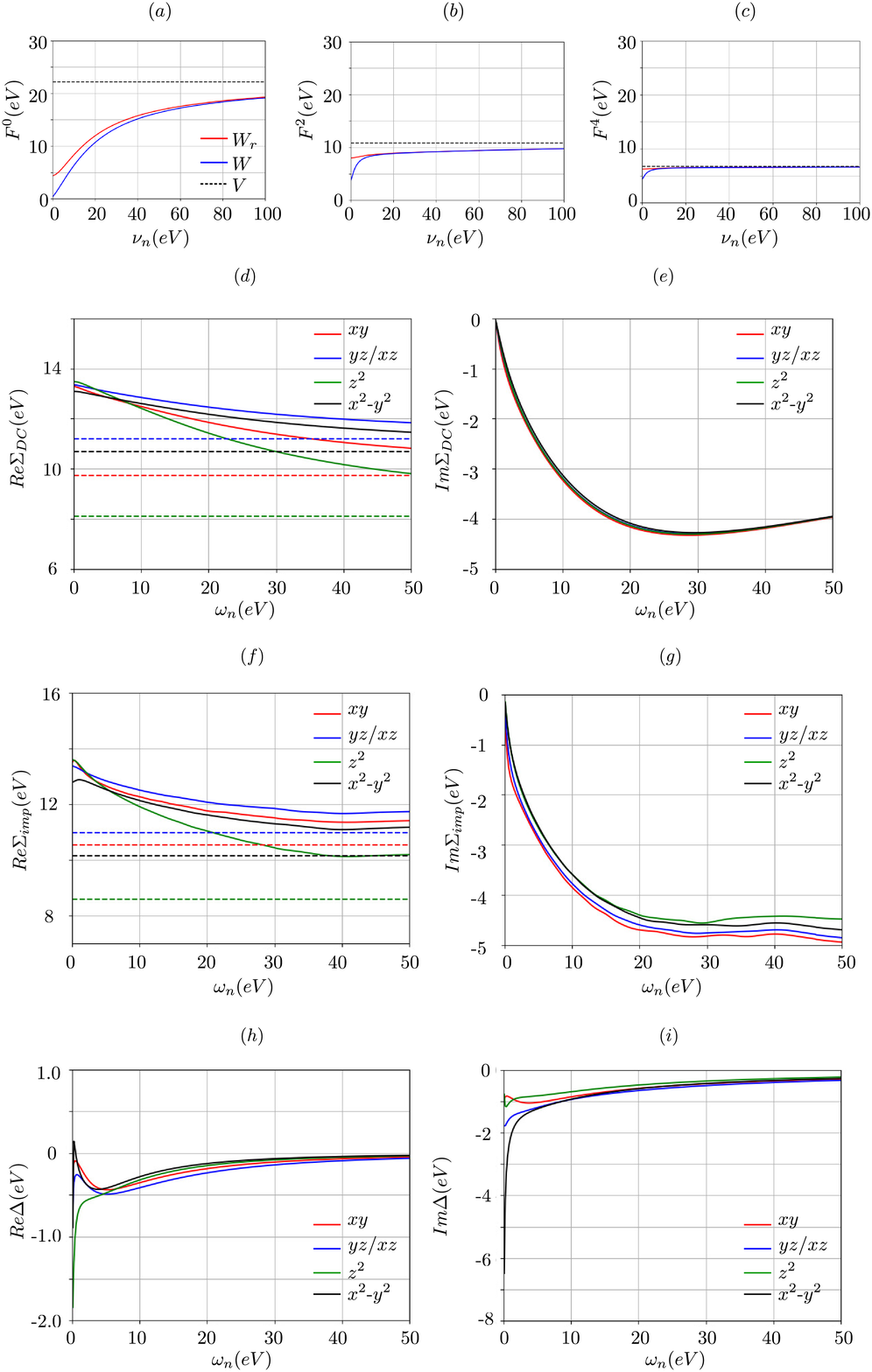}
  \caption{(color online) Slater's integrals associated with Fe-$d$ orbitals: (a) $F^0$, (b) $F^2$, and (c) $F^4$. Red and blue colored lines correspond to the Slater's integrals of partially screened Coulomb interaction $W_r$ and fully screened Coulomb interaction $W$, respectively. Black dashed lines are for bare Coulomb interaction. For DMFT calculation, only dynamical screening associated $F^0$ has been considered. (d) real and (e) imaginary parts of spin and orbital resolved local-GW self-energies associated with a Fe-$3d$ orbital. Dotted lines show their high-frequency limit. (e) real and (f) imaginary part of spin and orbital resolved DMFT self-energies associated with a Fe-$3d$ orbital. (g) real and (h) imaginary parts of spin and orbital resolved hybridization function associated with a Fe-$3d$ orbital.}
  \label{fig_fese_imp}
\end{figure}

\subsection{The density of states and spectral functions of FeSe}
In this subsection, total DOS, projected DOS and crystal momentum resolved spectral function of FeSe are presented. Five Fe-$d$ orbitals are considered as correlated orbitals and Wannier functions for Fe-$s$, Fe-$p$ Fe-$d$, Se-$p$ and Se-$d$ orbitals are constructed in a frozen energy window of $\text{-}10 eV <E\text{-}E_F < 10eV$. Physical quantities associated with Ni-$3d$ orbitals for the impurity problem are shown in fig. \ref{fig_fese_imp}. For comparison, charge self-consistent LDA+DMFT calculation is performed by using ComDMFT. For the LDA+DMFT, five Fe-$d$ orbitals are considered as correlated orbitals. To define five Fe-$d$ orbitals, Wannier functions for Fe-$s$, Fe-$p$ Fe-$d$, Se-$p$ and Se-$d$ orbitals are constructed in a frozen energy window of $\text{-}10 eV <E-E_f < 10eV$ at every charge self-consistency loop. $F^0=5.0eV$, $F^2=6.9eV$ and $F^4=4.3eV$ are chosen to construct Coulomb interaction tensor associated Fe-$d$ orbitals, corresponding to $U=5.0eV$ and $J=0.8eV$ \cite{kutepov_SelfconsistentGWDetermination_2010}. For the double-counting energy, nominal double-counting scheme \cite{haule_DynamicalMeanfieldTheory_2010,haule_CovalencyTransitionmetalOxides_2014} is used with $d$ orbital occupancy of 6.0. LDA+DMFT results from ComDMFT reproduce well earlier results \cite{mandal_StrongPressuredependentElectronphonon_2014,yin_SpinDynamicsOrbitalantiphase_2014, haule_ForcesStructuralOptimizations_2016,yin_KineticFrustrationNature_2011} from EDMFTF\cite{haule_DynamicalMeanfieldTheory_2010,_EmbeddedDMFTFunctional_}. Spin-orbit coupling is neglected in all methods. 

\begin{figure}
  \centering
  \includegraphics[width=1.0\columnwidth]{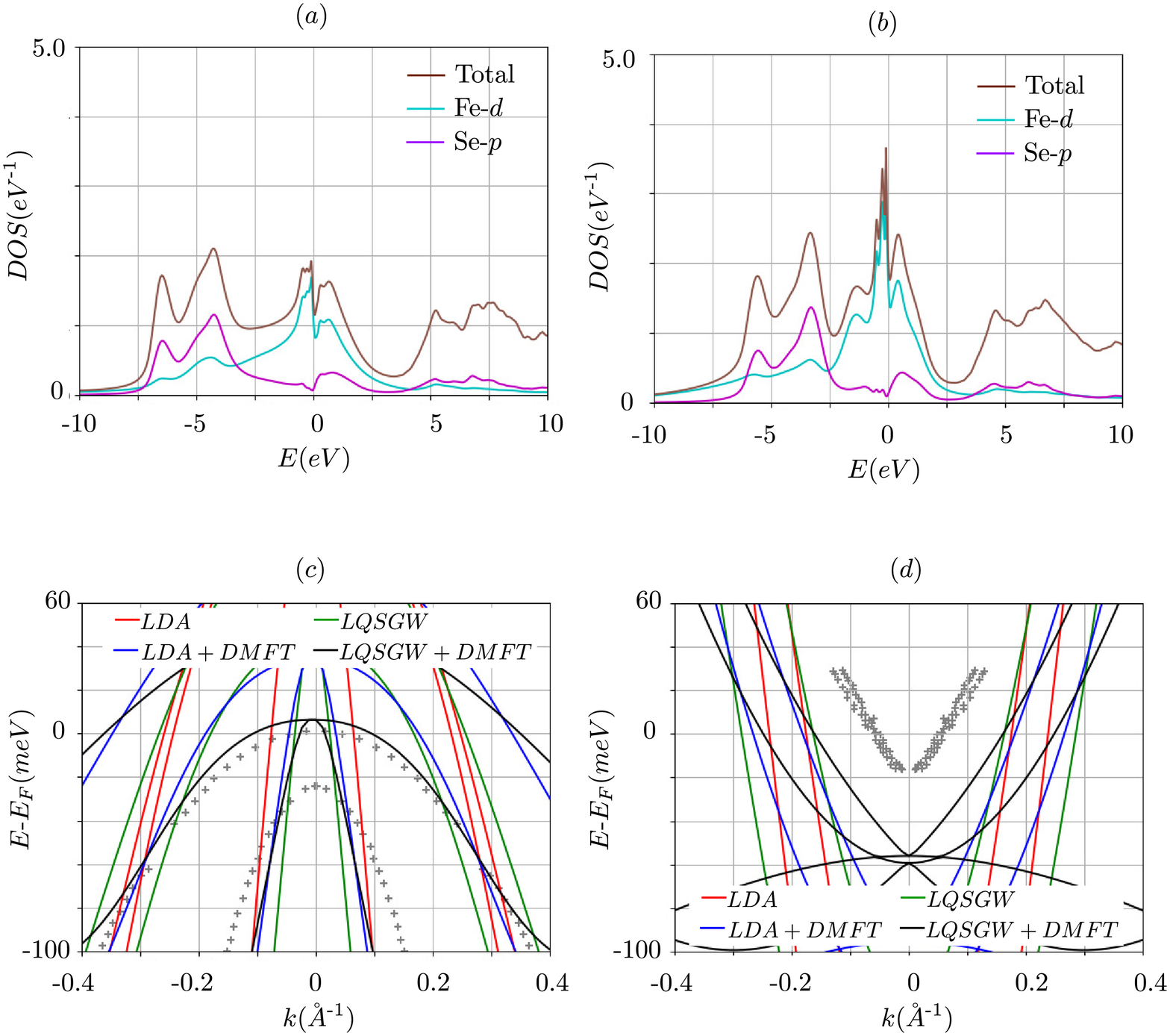}
  \caption{(color online) the total density of states and the projected density of states within (a) \textit{ab initio} LQSGW+DMFT and (b) charge self-consistent LDA+DMFT. FeSe bandstructure within LDA (red lines), LQSGW (green lines), LDA+DMFT (blue lines), and LQSGW+DMFT (black lines) along (c) $M$-$\Gamma$-$M$ line and (d) $\Gamma$-$M$-$\Gamma$ lines at T=300K. Angle-resolved photoemission data at high-temperature phase (T=120K) are marked by gray plus symbols \cite{watson_EmergenceNematicElectronic_2015,zhang_ObservationTwoDistinct_2015}}
  \label{fig_spectra_fese}
\end{figure}

Figure \ref{fig_spectra_fese} (a) and (b) show FeSe DOS within \textit{ab initio} LQSGW+DMFT and charge self-consistent LDA+DMFT, respectively. They show qualitatively similar line shapes, although the \textit{ab initio} LQSGW+DMFT lineshape is more extended in energy than charge self-consistent LDA+DMFT lineshape. Figure \ref{fig_spectra_fese} (c) and (d) show quasiparticle bandstructure along M-$\Gamma$-M line and $\Gamma$-M-$\Gamma$ line within four different theories: LDA (red lines), LQSGW (green lines), LDA+DMFT (blue lines), and LQSGW+DMFT (black lines). For the construction of the quasiparticle bands within LDA+DMFT and LQSGW+DMFT, we linearized impurity self-energy around the Fermi level and constructed quasiparticle Hamiltonian in the following way.

\begin{equation}
  \begin{split}
    H_{QP}(\mathbf{k})=\sqrt{Z_{imp}(\mathbf{k})} \left(H_{MF}(\mathbf{k})+f_\mathbf{k} \widetilde{\Sigma}_{imp}(\omega=0) f_\mathbf{k}^\dagger\right)\sqrt{Z_{imp}(\mathbf{k})},
  \end{split}
\end{equation}
where $Z_{imp}^{-1}(\mathbf{k})=1-f_\mathbf{k}\left({\partial{\widetilde{\Sigma}_{imp}}}/{\partial{i\omega_n}}|_{\omega=0}\right)f_\mathbf{k}^\dagger$. $H_{MF}$ is $H_{QP}^{nl}$ in eq. \eqref{eq:h_qp_nl} for LQSGW+DMFT and nominal-double-counting-term-corrected Kohm-Sham Hamiltonian ($H_{DFT}(\mathbf{k})-f_\mathbf{k} \widetilde{\Sigma}_{DC}f_\mathbf{k}^\dagger$) for charge self-consistent LDA+DMFT, respectively. For comparison, quasiparticle bands from angle-resolved photoemission spectroscopy are reproduced \cite{watson_EmergenceNematicElectronic_2015,zhang_ObservationTwoDistinct_2015}. Near $\Gamma$ point, in its high-temperature phase, a single hole packet has been observed experimentally and it is attributed to Fe-$d_{xz}$ and Fe-$d_{yz}$ orbitals \cite{coldea_KeyIngredientsElectronic_2018}. Due to spin-orbit coupling, the degeneracy at $\Gamma$ is lifted and another band below the Fermi level has been observed. Among four different \textit{ab initio} methods, LQSGW+DMFT shows the best agreement with the experiments in terms of the effective mass for Fe-$d_{xz}$ and Fe-$d_{yz}$ bands and the position of the bands at the $\Gamma$. Here, we note that all four different methods predict that there is another hole pocket from Fe-$d_{xy}$ orbitals at $\Gamma$ points which has not seen from ARPES experiments. Near $M$ point, it is expected to see two electron pockets: one from Fe-$d_{xz}$ and Fe-$d_{yz}$ orbitals and the other from Fe-$d_{xy}$. From experiments, only a single electron pocket has been observed yet. Among four different \textit{ab initio} methods, LQSGW+DMFT results in the best agreement with the experiments in terms of the effective mass for the electron bands and the position of the bands at the $M$.

\section{Summary}
We presented the implementation of \textit{ab initio} LQSGW+DMFT and charge self-consistent LDA+DMFT for the electronic structure of CES in ComDMFT. The rationale for this development is to provide multiple methods in one open-source GPL license platform in order to investigate the consequences of the coexistence of localized and itinerant characters of correlated-electrons and to potentiate a diagrammatically motivated \textit{ab initio} approaches for CES. This code will serve as a starting point for other \textit{ab initio} approaches on CES.

\section{Acknowledgments}
This work was supported by the U.S Department of Energy, Office of Science, Basic Energy Sciences as a part of the Computational Materials Science Program.





\bibliographystyle{elsarticle-num}
\bibliography{zotero}







\end{document}